\documentclass[a4paper,11pt]{article}

\usepackage{graphicx,array}
\usepackage{color}
\usepackage{latexsym}
\usepackage{amsthm}
\usepackage{amsmath}
\usepackage{amssymb}
\usepackage{siunitx}
\usepackage{empheq}
\usepackage{braket}
\usepackage[version=4]{mhchem}

\usepackage{dsfont}
\usepackage{epsfig}
\usepackage{slashed}
\usepackage{bbold}
\usepackage{psfrag}
\usepackage[svgnames]{xcolor}
\PassOptionsToPackage{caption=false}{subfig}
\usepackage{subcaption}
\usepackage{xfrac}
\usepackage{multirow}
\usepackage{booktabs}
\usepackage{cite}
\usepackage[normalem]{ulem}
\usepackage{jheppub} % for details on the use of the package, please see the JHEP-author-manual
\usepackage{adjustbox}

\newcommand{\be}{\begin{equation}}
\newcommand{\ee}{\end{equation}}
\newcommand{\bea}{\begin{eqnarray}}
\newcommand{\eea}{\end{eqnarray}}
\newcommand{\bei}{\begin{itemize}}
\newcommand{\eei}{\end{itemize}}

\newcommand{\nn}{\nonumber}
\newcommand{\M}{M_\text{BSM}}

\title{On the Atomki nuclear anomaly after the MEG-II result}

\author[a]{D. Barducci,}\emailAdd{daniele.barducci@pi.infn.it}
\author[b]{D. Germani,}\emailAdd{davide.germani@uniroma1.it}
\author[b]{M. Nardecchia,}\emailAdd{marco.nardecchia@roma1.infn.it}
\author[b]{S. Scacco,}\emailAdd{s.scacco@uniroma1.it}
\author[c,d]{C. Toni}\emailAdd{claudio.toni@lapth.cnrs.fr}

\affiliation[a]{
Dipartimento di Fisica Enrico Fermi, Universit\`a di Pisa and INFN, Sezione di Pisa,
Largo Bruno Pontecorvo 3, 56127 Pisa, Italy
}
\affiliation[b]{
Universit\`a degli Studi di Roma La Sapienza and INFN Section of Roma 1, Piazzale Aldo Moro 5, 00185 Roma, Italy
}
\affiliation[c]{
Dipartimento di Fisica e Astronomia `G.~Galilei', Universit\`a di Padova and INFN Sezione di Padova, Via F. Marzolo 8, 35131 Padova, Italy
}
\affiliation[d]{
LAPTh, Université Savoie Mont-Blanc et CNRS, 74941 Annecy, France
}

\abstract{
Recent experimental results from the Atomki collaboration have reported the observation of anomalous effects in Beryllium, Helium and Carbon nuclear transitions that could hint at physics beyond the Standard Model. However, the MEG-II experiment has recently found no significant anomalous signal in the Beryllium transition ${^8}\text{Be}^\star\to{^8}\text{Be}+e^+e^-$. In view of this result, we critically re-examine the possible theoretical interpretations of the anomalies observed by the Atomki experiment in terms of a new  boson $X$ with mass around $17\;$MeV.
The present work aims to study the phenomenology of a spin-2 state and revisit the possibility of a pure CP-even scalar, which was initially dismissed due to its inability to explain the Beryllium anomalous signal. Our analysis shows that a spin-2 state is highly disfavoured by the SINDRUM constraint while a scalar boson could explain the Helium and Carbon anomalies while being compatible with other experimental constraints.
}

\begin{document} 
\maketitle

\newpage

\section{Introduction}

Rare nuclear transitions are a powerful tool for observing New Physics (NP) degrees of freedom with mass at the MeV scale and weakly coupled to the Standard Model (SM) fields.
Nuclear transitions are the processes through which an
excited nucleus decays to a lower energy level and, within the SM, can only be mediated by electromagnetic (EM) interactions. They can mainly proceed through the following channels
\begin{itemize}
\item $\gamma-$emission, where the nucleus decays emitting a real photon,
\item Internal Pair Creation (IPC), where the nucleus emits a virtual photon which then decays to an $e^+e^-$ pair.
\end{itemize}

In the last decade the Atomki collaboration has reported various anomalous measurements in the IPC decays of excited $^8$Be\,\cite{Krasznahorkay:2015iga,Krasznahorkay:2018snd}, $^4$He\,\cite{Krasznahorkay:2019lyl,Krasznahorkay:2021joi} and $^{12}$C\,\cite{Krasznahorkay:2022pxs} nuclei. 
The observed anomalies appear as bumps for both the invariant mass and the angular opening of the $e^+e^-$ pairs and have 
a high statistical significance, well above $5\sigma$. Atomki has proposed to interpret them as the on-shell emission of a new boson $X$ from the excited nuclei, followed by its decay to an electron-positron pair.
A fit of the angular observables measured by the Atomki collaboration allows to estimate the mass of the hypothetical particle, which is found to be $m_X=(16.85\pm0.04)$ MeV \cite{Denton:2023gat}.

For a long time, no independent confirmation of these anomalies has been provided\footnote{With the exception of an experiment conducted at Vietnam National University \cite{Anh:2024req} which seems to confirm the Beryllium anomaly.},
despite an increasing interest of the high-energy community \cite{Feng:2016jff,Feng:2016ysn,Kozaczuk:2016nma,Ellwanger:2016wfe,Zhang:2017zap,Alves:2017avw,DelleRose:2018eic,Pulice:2019xel,Feng:2020mbt,Zhang:2020ukq,Alves:2020xhf,Hostert:2020xku,Chen:2020arr,Barducci:2022lqd,Kubarovsky:2022zxm,Wong:2022kyg,Viviani:2021stx,Alves:2023ree,Gysbers:2023wug,Denton:2023gat,Hostert:2023tkg,Mommers:2024qzy} and many experimental proposals\,\cite{MEGII:2018kmf,Chiappini:2022egy,Azuelos:2022nbu,Cortez:2023ycv,Raggi:2014zpa,Raggi:2015gza,Nardi:2018cxi,Darme:2022zfw,Mommers:2024ugr}.
Among these proposals, the MEG-II apparatus at PSI stands out. Although it was originally designed to search for the $\mu^+\to e^+\gamma$ decay, it also has the capacity to measure the very same Beryllium transition investigated by Atomki.
Recently, the MEG-II collaboration has released the results of an analysis with a four-week data-taking in 2023\,\cite{MEGII:2024urz}. 
MEG-II observes no significant signal above the expected background, presenting a stark contrast to the Atomki results. Nevertheless, the Beryllium anomaly cannot be definitively excluded, as MEG-II reports its result to be compatible with the Atomki ones within $1.5\sigma$. Conclusively resolving this tension  will require further experimental investigation. In this regard, MEG-II plans to collect additional data, which will either strengthen their null result or potentially reveal an anomalous signal.

While awaiting for more updates to shed light on this puzzle, this work aims to review the possible interpretation of the experimental data in terms of a new BSM boson $X$, in light of the recent MEG-II results.

Compared to a previous analysis performed by two authors of this work\,\cite{Barducci:2022lqd}, we enlarge the possible spin-assignments of the hypothetical boson, by also studying the phenomenology of a
spin-2 state with both positive and negative parity.
Furthermore, we reconsider the possibility of a CP-even scalar, an hypothesis that was initially discarded due to the incompatibility with the anomalous signal observed in Beryllium transition under the assumption of parity conservation. A CP-even scalar 
could serve as 
potential solution to the anomalies observed in the Helium and Carbon data and that will become relevant in case the null result from the MEG-II search in Beryllium transitions will be confirmed.

For our analysis, we employ a multipole expansion method and
give an estimate for the range of values of the effective nucleon couplings to the new light state in order to match the experimental observations.
We will show that a spin-2 state is highly disfavoured by the SINDRUM search
for the decay $\pi^+ \to e^+ \nu_e X$\,\cite{SINDRUM:1986klz},
while a CP-even scalar boson could potentially accommodate the Helium and Carbon anomalies while being compatible with other experimental bounds.

The paper is structured as follows.
In Sec.\,\ref{sec:review} we review the anomalous measurements performed by the Atomki experiment, the null result from MEG-II and comment on the possible theoretical interpretations.
In Sec.\,\ref{sec_sig} we describe the multipole expansion formalism while in Sec.\,\ref{sec_X17_0} and Sec.\,\ref{sec_X17_2} we detail the calculations for the decay rates of the $^8$Be, $^4$He and $^{12}$C resonances for the spin-parity assignments considered in this work.
We present our results in Sec.\,\ref{sec:res} and we then conclude in Sec.~\ref{sec:conc}. We also add some appendices with more technical details.
In App.\,\ref{App:non_rel_spin2} we derive the nonrelativistic expansion of the nuclear tensor operators 
and in App.\,\ref{App:chiSQ} we present some details of the stastycal procedure that we have adopeted.
In App.\,\ref{App:sindrum} we calculate the rate of the charged pion decay $\pi^+\to e^+ \nu_e X$ for a scalar and spin-2 boson. Finally in App.\,\ref{App:bound0} we summaryze the relevant bounds for a scalar boson.

\section{The Atomki anomaly after the MEG-II result}
\label{sec:review}

\begin{table}[t!]
\begin{center}
\begin{tabular}{c|c|c|c|c}
$N$ & $N^\star$ & $J^{\pi}$ & $I$ & $\Gamma$ [keV] \\
\midrule
\midrule
$^{8}\text{Be}$ &  & $0^{+}$ & 0 & $5.57\pm0.25$ \\
\midrule
& $^{8}\text{Be}(18.15)$ & $1^{+}$ & $0^{*}$ & $138\pm6$ \\
\midrule\midrule
$^{4}\text{He}$ & & $0^{+}$ & $0$ & $\text{Stable}$ \\
\midrule
& $^{4}\text{He}(21.01)$ & $0^{-}$ & $0$ & $0.84$ \\
\midrule
& $^{4}\text{He}(20.21)$ & $0^{+}$ & $0$ & $0.50$ \\
\midrule\midrule
$^{12}\text{C}$ & & $0^{+}$ & 0 & Stable \\
\midrule
& $^{12}\text{C}(17.23)$ & $1^{-}$ & $1$ & $1150$ \\
\end{tabular}
\end{center}
\caption{Spin-parity $J^{\pi}$ and isospin $I$ quantum numbers, total decay widths for the nuclei used in the Atomki experiment: 
$^{8}\text{Be}$ \cite{Tilley:2004zz}, $^{4}$He \cite{Tilley:1992zz,Walcher:1970vkv} and $^{12}$C \cite{Kelley:2017qgh,Segel:1965zz} nuclei. Asterisks on isospin assignments indicate states with significant isospin mixing.}
\label{tab:nuclei}
\end{table}

In the Atomki experiment\,\cite{Gulyas:2015mia} a proton beam collides with a target nucleus $A$ at rest, with the aim of producing an excited nucleus $N^{*}$ and measure its IPC transition to a ground state $N$, \emph{i.e.}
\be
p+A \rightarrow N^\star \rightarrow N + e^+ e^- \ .
\ee
In Tab.~\ref{tab:nuclei}, we present all the ground and excited states investigated in the Atomki search, along with their quantum numbers and decay widths.
In each experiment, Atomki observed an anomalous peaks of events compatible with a  boson with mass around 17 MeV emitted in the transition and decaying to an electron-positron pair.
By employing the narrow-width approximation for the nuclear excited state, the main observable measured by Atomki in the Beryllium and Carbon experiments is the anomalous decay rate normalized to the $\gamma$-emission rate, \emph{i.e.} the decay ratios
\be
R_N\equiv\frac{\Gamma(N^\star\to N +X)}{\Gamma(N^\star\to N+\gamma)}\text{BR}(X\to e^+e^-)
\quad \text{for $N=\text{Be, C}$\, .}
\ee
Meanwhile, in the Helium experiment two excited states are populated, whose $\gamma$-emission is forbidden, and both of them could contribute to the anomalous signal. Thus, the decay ratio is defined in the Helium case as
\be
\begin{split}
R_\text{He}\equiv&\frac{\sigma(p+{^{3}\text{H}}\to {^4 \text{He}(20.49)} + X)}{\sigma(p+{^{3}\text{H}}\to {^4 \text{He}(20.49)} + e^+e^-)}\text{BR}(X\to e^+e^-) \\
=&\frac{\Gamma({^4 \text{He}}(20.21) \to {^4 \text{He}} \, + \, X)}{\Gamma({^4 \text{He}}(20.21) \to {^4 \text{He}} \, + \, e^{+}e^{-})}\text{BR}(X\to e^+e^-) \\
+&\frac{\sigma_{-}\Gamma_{+}}{\sigma_{+}\Gamma_{-}}\frac{\Gamma({^4 \text{He}}(21.01) \to {^4 \text{He}} \, + \, X)}{\Gamma({^4 \text{He}}(20.21) \to {^4 \text{He}} \, + \, e^{+}e^{-})} \text{BR}(X\to e^+e^-) \ ,
\end{split}
\ee
where $\Gamma_{\pm}$ are the total widths of the $0^{\pm}$ excited states of Helium nucleus while 
\begin{gather}
\sigma_{+}=\sigma(p+{^{3}\text{H}}\to {^4 \text{He}}(20.21))\quad \text{and} \quad \sigma_{-}=\sigma(p+{^{3}\text{H}}\to {^4 \text{He}}(21.01)) \ .
\end{gather}
The ratio $\sigma_{-}/\sigma_{+}$ can be evaluated as in App.~E of Ref.\,\cite{Barducci:2022lqd} in the narrow-width approximation.
\begin{table}[t!]
\begin{center}
\begin{tabular}{c|c|c|c}
 & $R_\text{Be}$ [$10^{-6}$] & $R_\text{He}$ & $R_\text{C}$ [$10^{-6}$] \\
\midrule
\midrule
Atomki & $6\pm1$\,\cite{Krasznahorkay:2015iga,Krasznahorkay:2018snd} & $0.2\pm0.03$\,\cite{Krasznahorkay:2019lyl,Krasznahorkay:2021joi} & $3.6\pm0.3$\,\cite{Krasznahorkay:2022pxs}\\
\midrule
MEG-II & $<5.3$ at 90\% CL\,\cite{MEGII:2024urz} &  &  \\
\midrule
Combined & $5.5\pm1.0$ &  &  \\
\end{tabular}
\end{center}
\caption{Experimental values at $1\sigma$ for the normalized decays rates from Atomki and the upper limit from MEG-II considering a mass value of $16.85$\,MeV for the $X$ boson. We associate to the Atomki Helium experimental value a relative uncertainty equal to the Beryllium one, as this information is not reported by Atomki. The Atomki and MEG-II results in the Beryllium case are combined as explained in the text.}
\label{tab:exp_value}
\end{table}

\begin{figure}[t!]
\begin{center}
\includegraphics[scale=0.5]{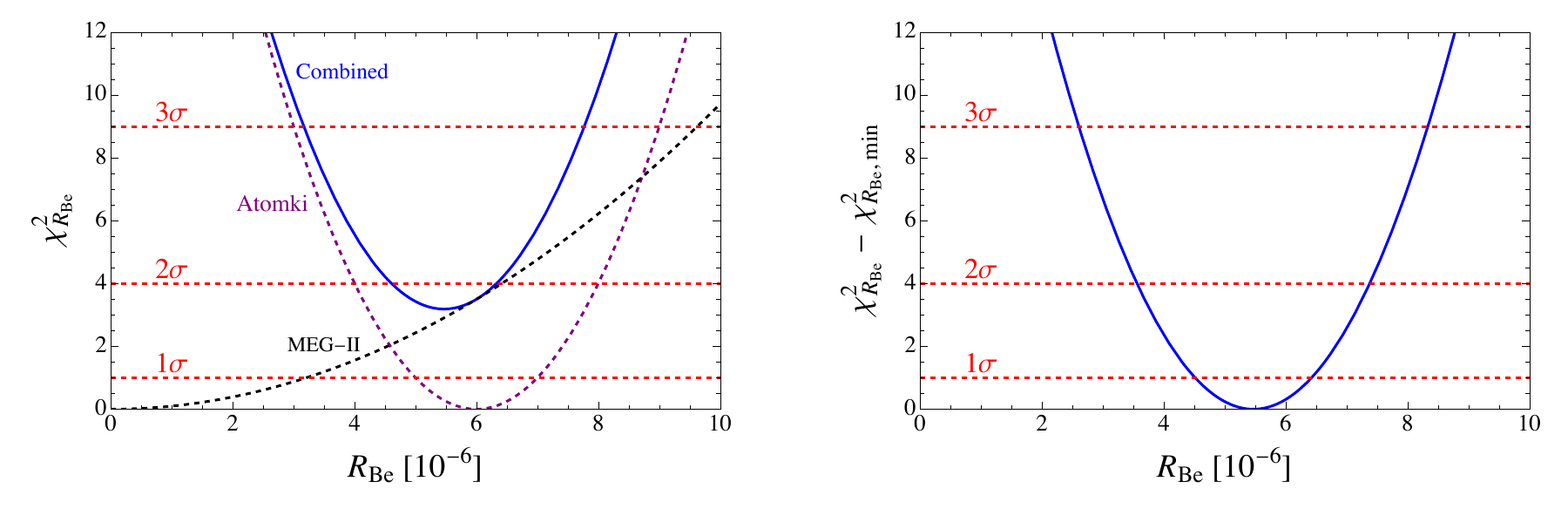}
\caption{\emph{Left panel:} Value of the $\chi^2$ of the Atomki (purple dashed line) and MEG-II (black dashed line) measurements and the sum of them (blue line). The red dashed lines denotes the 68\%, 95\%, 99\% values of a $\chi^2$ variable with 1 degree of freedom. \emph{Right panel:} Value of the total $\chi^2$ around its minimal value.}
\label{fig:R_Be}
\end{center}
\end{figure}

We report the Atomki and MEG-II results in Tab.\,\ref{tab:exp_value}, while fixing $m_X=16.85$ MeV as in Ref.~\cite{Denton:2023gat}.
To combine the Atomki and MEG-II result in the Beryllium case, we construct a $\chi^2$ variable by summing the two contributions from both measurements.
The individual $\chi^2$ and their combination are shown in the left panel of
Fig.\,\ref{fig:R_Be}. Note that the minimum of the combined $\chi^2$ is within $2\sigma$ from both the measurements. Thus, no definitive exclusion of the Beryllium anomaly is reached. We therefore extract 
a best-fit value and a 
 68\%\,confidence level (CL) interval for the combined measurement, by rescaling the total $\chi^2$ to its minimun, as shown in the right panel of Fig.~\ref{fig:R_Be}. The results of the fit,
also reported in Tab.\,\ref{tab:exp_value}, 
 yields $R_\text{Be}=(5.5\pm1.0)\times10^{-6}$.

In light of the recent result by the MEG-II collaboration and our statistical interpretation, we critically re-examine the possible theoretical interpretations of the anomaly in terms of a new BSM state, updating the picture exposed in Ref.~\cite{Barducci:2022lqd}. A pseudoscalar, $S^\pi=0^-$, or a vector, $S^\pi=1^-$, solution remains disfavored, as the former cannot be emitted in the Carbon transition, while the latter is well constrained by the NA48 experiment \cite{NA482:2015wmo} and, as recently pointed out by Ref.\,\cite{Hostert:2023tkg}, SINDRUM searches \cite{SINDRUM:1986klz}. Regarding the axial vector, $S^\pi=1^+$, solution, which has been identified as the most promising in\,\cite{Barducci:2022lqd}, recently Ref.\,\cite{Mommers:2024qzy} computed the missing axial nuclear matrix element of the Carbon transition in a hole-particle approximation,
for which\,\cite{Barducci:2022lqd} only made an order of magnitude estimate. With this result the theoretical prediction in terms of an $S^\pi=1^+$ boson is in tension with the SINDRUM constraint.

In this work, we then broaden the possible spin-parity assignments by investigating the phenomenology of a spin-2 state with both positive and negative parity and reconsider the possibility of a CP even scalar.
In addition to the electron-positron channel, both a spin-2 and scalar states can also decay into a pair photons, producing a different experimental signature. In 2019, Atomki searched for a $2\gamma$ final state measuring the nuclear process ${^3}\text{H}(n,\gamma\gamma){^4}\text{He}$ but no significant signal was found over the considerable cosmic ray background \cite{Nagy:2019pdt}. More recently, the JINR collaboration has reported anomalous peaks of di-photon events produced in nuclear reactions, with invariant masses around $17$ and $38$ MeV\,\cite{Abraamyan:2023hed}, where the former can be potentially related with the Atomki anomaly. Although we do not perform an assessment of the compatibility with the results of\,\cite{Abraamyan:2023hed}, we will also consider the possibility of a photon coupling\footnote{Instead we neglect a coupling to neutrinos for simplicity.} for the $X$ boson, and briefly comment on the possible relation with the JINR peak.

\section{Signal computation: overview}
\label{sec_sig}

The new physics required to explain the potential anomalous results implies couplings of the new state with quarks and leptons. From the fundamental physics Lagrangian, an effective Hamiltonian for nucleons can be derived. In what follows, the only relevant information is the assignment of the particle's spin and parity quantum numbers.
In this section we introduce the formalism used in order to estimate the anomalous nuclear decay rates relevant for the Atomki experiments, see also~\cite{Kozaczuk:2016nma,Barducci:2022lqd}.
At low energy, we  describe the nuclear interaction of a spin-$s$ boson through the effective Hamiltonian
\be
H_{\rm int}^{s}=\begin{cases}
\int\!d^{3}\vec{r}\,\mathcal{S}\!(\vec{r}) X\!(\vec{r}) &\quad \text{if $s=0$ ,}\\
\int\!d^{3}\vec{r} \ \mathcal{H}_{\mu\nu}\!(\vec{r}) X^{\mu\nu}\!(\vec{r}) &\quad  \text{if $s=2$ ,}
\end{cases}
\ee
where the nuclear scalar density $\mathcal{S}$ and the nuclear symmetric tensor $\mathcal{H}_{\mu\nu}$
encodes the quantum operators containing all the information of the nuclear matter fields.
At the lowest order in the interaction picture, the nuclear matrix element is given by
\be
\label{eq:matrix-element}
\mathcal{T}_{fi}^{s}=
\braket{N,X|H_{\rm int}^{s}|N^\star}=\begin{cases}
\bra{N}\int\!d^{3}\vec{r}\,\mathcal{S}\!(\vec{r}) e^{-i\vec{k}\cdot\vec{r}} \ket{N^\star} & \text{if $s=0$ ,}\\
\bra{N}\int\!d^{3}\vec{r}\,[\epsilon_{a}^{\mu\nu}\!(\vec{k})]^{*}\mathcal{H}_{\mu\nu}\!(\vec{r}) e^{-i\vec{k}\cdot\vec{r}} \ket{N^\star} & \text{if $s=2$ ,}
\end{cases}
\ee
where $\vec k$ is the boson momentum, the index $a=0,\pm1,\pm2$ labels the polarizations of a spin-2 boson while $\ket{N^\star}=\ket{J_{i}M_{i}}$ and $\ket{N}=\ket{J_{f}M_{f}}$ indicate the nuclear matter initial and final states.

\subsection{Multipole expansion}

To compute the decay widths of the nuclear transition, it is useful to expand the nuclear matrix elements in terms of spherical tensor operators ${\cal O}_{JM}$ involving $\mathcal{S}$ or $\mathcal{H}_{\mu\nu}$. Symbolically,
\be
\mathcal{T}_{fi}^{s}=\bra{N} \sum_{{\cal O}}\sum_{JM}
{\cal O}_{JM} \ket{N^\star} \ .
\ee
The reason is that spherical operators satisfy the Wigner-Eckart theorem \cite{book:17167}
\be
\begin{split}\label{eq:wigner}
\braket{J_{f}M_{f}|{\cal O}_{J,-M}|J_{i}M_{i}}=\frac{(-1)^{J_{i}-M_{i}}}{\sqrt{2J+1}}C^{J,-M}_{J_f, M_f; J_i, -M_i}\braket{J_{f}||{\cal O}_{J}||J_{i}} \ ,
\end{split}
\ee
where the reduced matrix element $\braket{J_{f}||{\cal O}_{J}||J_{i}}$ contains all the physical information of the operator while its behavior under rotation is completely set by the Clebsh-Gordan coefficient  $C^{J,-M}_{J_f, M_f; J_i, -M_i}$.
For a spin-0 boson, one gets~\cite{Barducci:2022lqd,book:80102}
\be
\begin{split}
\Gamma(N^\star\to N X)
&=\frac{1}{2J_{i}+1}\frac{k}{2\pi}\sum_{\text{pol.}}\left|\mathcal{T}_{fi}^{s=0}\right|^2 \\
&=\frac{2k}{2J_{i}+1}\sum_{J\geq 0}\left|\braket{N||\mathcal{G}_{J}||N^\star}\right|^{2} \ ,
\end{split}
\ee
with the spherical operators defined as
\be
\label{eq:Gop}
\mathcal{G}_{JM} =
\int d^{3}\vec{r} \ \, j_{J}(kr)Y_{JM}(\hat{r})\mathcal{S}(\vec{r}) \ ,
\ee
where $k=|\vec{k}|$, $r=|\vec{r}|$ and $\hat{r}=\vec{r}/r$ while $Y_{JM}$ and $j_{J}(x)$ are respectively the 
spherical harmonics and the spherical Bessel functions, see {\it e.g.}\,\cite{book:17167}.

The angular momentum conservation law, encoded in the Clebsh-Gordan coefficients of Eq.~\eqref{eq:wigner}, states that the matrix element of the spherical operators vanishes, unless the following conditions are met
\begin{align}
& |J_{f}-J_{i}|\leq J\leq J_{f}+J_{i}\ , \nonumber \\
& M=M_{i}-M_{f} \ .
\end{align}
Moreover, if the $X$ boson has a definite parity $\pi_X$,  additional constraints on the matrix elements come from the requirement of parity conservation, \emph{i.e.} $\braket{N||\mathcal{O}_{J}||N^\star}=0$ unless $\pi_{N^\star}=\pi_{N}\pi_{{\cal O}}$.
The reduced matrix elements are then evaluated in the limit of the nonrelativistic and long wavelenght expansion, which we describe below.

\subsection{Nonrelativistic expansion for nuclear operators} \label{nonrel}

Statistical arguments\,\cite{book:14848} allow to estimate 
the maximal kinetic energy $E_{\rm c}$ per nucleon in the nucleus to be of the order of 30\;MeV. This implies that 
a nucleus can be modeled as a quantum mechanical system of nonrelativistic point-like nucleons~\cite{book:80102}, enabling to take
the nonrelativistic limit of the nuclear operators and to write them in first quantization formalism. Any nuclear operator is given by
\be
\mathcal{O}(\vec{r})=\sum_{s=1}^{A}\hat{\mathcal{O}}_{s}^{(1)}(\vec{r}-\vec{r}_{s}) \ ,
\label{eq:nuc_op}
\ee
with the single particle operator $\hat{\mathcal{O}}_{s}^{(1)}(\vec{r}-\vec{r}_{s})\propto\delta(\vec{r}-\vec{r}_{s})$ in the nucleon point-like approximation. We want to match the expression of Eq.~\eqref{eq:nuc_op} 
with its relativistic counterpart, where the nucleons are described in terms of quantum fields $p(x)$ and $n(x)$ and the nucleon operators we will consider in this work are bilinears in $p(x)$ and $n(x)$.
For operators which are even or odd under parity, higher order terms in the expansions are of order of $\braket{p_N}^2/m_N^2\sim 6\times 10^{-2}$ with respect to the leading one, where $\braket{p_N}$ is the average nucleon momentum, and can then be safely neglected.

\subsection{Long wavelength approximation}

The nuclear radius is approximately given by $R \simeq R_0 A^{\frac{1}{3}} \simeq 6.1 \times 10^{-3} A^{\frac{1}{3}}\;$MeV$^{-1}$ \cite{book:80102}, which implies that, in all the cases of interest, the nucleus size is significantly smaller than the boson wavelength $k^{-1}\sim (10\;{\rm MeV})^{-1}$.
Thus, one expands the spherical Bessel function in Eq.~\eqref{eq:Gop} for small $k r$ as
\begin{equation}
\label{lwl}
j_{J}(kr)\simeq\frac{(kr)^{J}}{(2J+1)!!} \ ,
\end{equation}
with higher order corrections contributing at order $(kr)^2\simeq 1\%$ with respect to the leading one for the cases of interest, which can therefore be neglected.

For a spin-2, we will directly expand the exponential factor in Eq.~\eqref{eq:matrix-element} as
\be
\label{eq:lwlexp}
e^{-i\vec{k}\cdot\vec{r}}\approx 1-i\vec{k}\cdot\vec{r}-\frac{1}{2}(\vec{k}\cdot\vec{r})^2+\dots \ .
\ee
Then, the product of the $n$-th power of $\vec{r}$ and the components of nuclear tensor operator ${\cal H}^{\mu\nu}$ can be decomposed into a sum of irreducible spherical operators. Parity conservation ensures that higher order corrections are again of order $(kr)^2\simeq 1\%$ with respect to the leading term. Thus, keeping the reduced nuclear matrix element of the leading spherical operators in the long wavelength and nonrelativistic expansions is sufficient to our scope.
The techniques introduced in this section will be implemented in the next one in order to compute the decay rates of the nuclear transitions.

\section{Signal computation: scalar boson dynamics}
\label{sec_X17_0}

In this section, we consider the case of a CP even scalar boson.
To describe its dynamics, we rely on an effective field theory (EFT) approach and consider the lowest-dimensional interacting operators. At renormalizable level, the interaction Lagrangian is given by\footnote{In principle, the effective nucleon coefficients are form factors $F(q^2)$. However, the transferred momentum is much smaller than the hadron scale $\Lambda_{\rm QCD}$. For all intents and purposes, form factors can be treated as constants here and in the following.}
\be
\label{eq:nuc_int_0}
{\cal L}_\text{int}^{d\leq4}= z_{p}\overline{p}pX  +z_{n}\overline{n}nX + z_e \overline{e}eX  \ ,
\ee
while a coupling to photons only appears at dimension five
\be
\label{eq:nuc_int_0_g}
{\cal L}_\text{int}^{d=5}=
\frac{\alpha}{8\pi}\frac{X}{f_\gamma}
F_{\mu\nu}F^{\mu\nu}  \ ,
\ee
with $\alpha$ the electromagnetic coupling and $f_\gamma$ the scale at which this effective operator is generated.
The suppression factor in front of the operator is explicitly introduced since this only arises at loop level in any weakly coupled UV completion with heavy charged particles. The masses of the latters give the EFT scale $f_\gamma$ and must lie above the electroweak scale to avoid the constraints on exotic charged particle direct searches. Hence, we reasonably expect this operator to have negligible contributions and we will not consider its effect in the following.
The matching between the effective nucleon interactions and the fundamental interactions of the $X$ boson with quark and gluons is instead reported in App.~C of Ref.~\cite{Barducci:2022lqd}.
In the long wavelength and nonrelativistic expansions, the nucleon terms in Eq.~\eqref{eq:nuc_int_0} yield\,\cite{Barducci:2022lqd}\footnote{The vectors $\hat e_M = \sqrt{4\pi/3} Y_{1M}(\hat r)$ form a tridimensional orthonormal basis and transform as $\ket{J=1,M}$ state.}
\begin{align}
\mathcal{G}_{00}\simeq & \frac{1}{\sqrt{4\pi}}
\sum_{s=1}^{A} z_{s}
\left[ 1 - \frac{p_{s}^{2}}{2m_{N}^{2}} - \frac{k^{2}r_{s}^{2}}{6} \right] \ , \\
\mathcal{G}_{1M}\simeq & \frac{k}{3}\sqrt{\frac{3}{4\pi}}
\sum_{s=1}^{A}z_{s}
\vec{r}_{s}\cdot \hat{e}_{M} \ ,
\end{align}
where $m_N$ is the nucleon mass while $z_{s}=z_{p}$ ($z_{s}=z_{n}$) if the $s$-th nucleon is a proton (neutron). A similar notation is adopted for the rest of the work. Note that the next-to-leading terms of ${\cal G}_{00}$ are needed because the leading term cannot mediate transitions among orthogonal states with zero isospin as in the case of the Helium transition.
The decay ratios are then given by
\begin{align}
R_\text{He}=&\frac{1}{\alpha^2} \frac{15}{8} \left(\frac{k}{\omega}\right)^5 (z_p+z_n)^2\left|1+3r_\text{He}\right|^2 \text{BR}(X\to e^+e^-) \ , \\
R_\text{C}=& \left(\frac{k}{\omega}\right)^3 \frac{(z_p-z_n)^2}{8\pi\alpha_e} \text{BR}(X\to e^+e^-) \ , 
\end{align}
with $\alpha$ the electromagnetic coupling, $\omega$ the energy released in the nuclear process, $k=\sqrt{\omega^2-m_X^2}$ and where we introduce
\be
r_\text{He}=\frac{\braket{\ce{^{4}He}|| \sum_{s=1}^A (p_s^2 / m_N^2) ||\ce{^{4}He}(20.21)}}{\braket{\ce{^{4}He}|| \sum_{s=1}^A (k^2 r_s^2) ||\ce{^{4}He}(20.21)}} \ .
\ee
Note that the Helium (Carbon) ratio depends only on the isoscalar (isovector) coupling due to the isospin symmetry.

\section{Signal computation: spin-2 boson dynamics}
\label{sec_X17_2}

In this section we discuss the case of a spin-2 particle. 
We assume this state to be composite rather than fundamental, as no renormalizable and complete model is possible in quantum field theory (QFT) for a spin-2 state.
More specifically, we propose the existence of a BSM sector which, below a composite scale denoted as $\Lambda_c$, includes in the spectrum a composite spin-2 state of mass around 17 MeV. Naively, we expect $\Lambda_c\sim4\pi m_X$. Additionally, we assume that the BSM sector interacts with the SM fields via heavy mediators at a scale $\M$, taken to be independent and much greater than the composite scale.
In QFT, spin-2 particles are described by a symmetric tensor field $X_{\mu\nu}$ with sixteen degrees of freedom, of which only five are physical. To avoid complications arising from dealing with a large number of unphysical degrees of freeedom, it is more convenient to take as a starting point the on-shell 3-point amplitudes of the $X$ boson, relevant for its decays and the nuclear transitions.

Following \cite{Panico:2016ary}, we parametrize the most general on-shell 3-point amplitude for the spin-2 state interaction to fermions, like nucleons or electrons, as 
\be
\label{eq:nuc_int}
\begin{split}
{\cal A}(f\to f^\prime X) = \overline{u}(p^\prime,\sigma^\prime)
\Biggr\{ &C_f \left[ \gamma_\mu \left(\frac{p^\prime+p}{4}\right)_\nu
+\gamma_\nu \left(\frac{p^\prime+p}{4}\right)_\mu
\right] \\
+&\tilde{C}_{f} \left[ \gamma_\mu \gamma_5\left(\frac{p^\prime+p}{4}\right)_\nu
+\gamma_\nu \gamma_5\left(\frac{p^\prime+p}{4}\right)_\mu
\right] \\
+& D_{f} \left(p^\prime+p\right)_\mu \left(p^\prime+p\right)_\nu \\
+& \tilde{D}_{f} \left(p^\prime+p\right)_\mu \left(p^\prime+p\right)_\nu i\gamma_5  \Biggr\} u(p,\sigma) \ [\epsilon_{a}^{\mu\nu}(p-p^\prime)]^* \ ,
\end{split}
\ee
with $p^{(\prime)}$ and $\sigma^{(\prime)}$ the momentum and spin of the fermions and
where we expect that the dimensional effective couplings we introduced parametrizing the CP-even and CP-odd interactions (the latter denoted with a tilded character) are of order of
\be\label{eq:spin2_a}
C_f\sim\tilde{C}_f\sim{\cal O}(\M^{-1})
\quad \text{and} \quad
D_f\sim\tilde{D}_f\sim{\cal O}(\M^{-2}) \ .
\ee
Since $\M$ is assumed to be much larger than the energy of the nuclear processes, the terms in the last two rows can be safely discarded.
Simirarly, we write the most general on-shell amplitude describing the di-photon decay of the spin-2 as \cite{Panico:2016ary}\footnote{We remark that the Schouten identity is used to eliminate all CP-odd structures in which $\epsilon_a^{\mu\nu}$ is contracted with $\epsilon_{1,2}^*$ or with the Levi-Civita tensor \cite{Panico:2016ary}.}
\begin{align}
{\cal A}(X\to \gamma\gamma)= 2i \Big\{ &C_\gamma\left[(\epsilon_1^* \cdot \epsilon_2^*) q_{1\mu} q_{2\nu} - (\epsilon_2^*\cdot q_1) \epsilon_{1\mu}^{*} q_{2\nu} - (\epsilon_1^*\cdot q_2) \epsilon_{2\mu}^* q_{1\nu} + (q_1 \cdot q_2) \epsilon_{1\mu}^{*} \epsilon_{2\nu}^{*}\right] \nn \\
\label{eq:gamma_int}
+& D_\gamma \, q_{1\mu} q_{2\nu} \left[(\epsilon_1^* \cdot \epsilon_2^*)(q_1 \cdot q_2) - (\epsilon_1^* \cdot k_2)(\epsilon_2^* \cdot k_1)\right] \\
+& \tilde{D}_\gamma \, q_{1\mu} q_{2\nu} \:\varepsilon^{\alpha\rho\beta\sigma}\,q_{1\alpha} q_{2\beta} \epsilon_{1\rho}^* \epsilon_{2\sigma}^*\Big\} \epsilon_{a}^{\mu\nu}(q_1+q_2) \nn\ ,
\end{align}
with $\epsilon_{1,2}$ and $q_{1,2}$ the polarization vectors and the momenta of the photons. The effective couplings
we introduced scale as
\be
C_\gamma\sim{\cal O}(\M^{-1})
\quad \text{and} \quad
D_\gamma\sim\tilde{D}_\gamma\sim{\cal O}(\M^{-3}) \ .
\ee
In the same spirit of Eq.\,\eqref{eq:spin2_a}, we will neglect the contribution of the last two couplings in the following.

In summary, we are left with seven couplings which are 
$C_{f=p,n,e}$ and $C_{\gamma}$ (only for $S^\pi=2^+$)
and
$\tilde{C}_{f=p,n,e}$ (only for $S^\pi=2^-$).

\subsection{Tensor boson $S^\pi=2^+$}

If the boson has positive parity, one has $\tilde{C}_f=0$.
From Eq.~\eqref{eq:nuc_int}, we derive the expressions of the nuclear tensor ${\cal H}^{\mu\nu}$ components at the leading order in the nonrelativistic limit, see App.~\ref{App:non_rel_spin2} for details.
Employing the long wavelength expansion as in Eq.~\eqref{eq:lwlexp}, we derive the decay ratios for the nuclear transitions of a spin-2 boson in terms of reduced matrix elements of spherical operators.
We remark that the interacting term proportional to the coupling $C_f$ in Eq.~\eqref{eq:nuc_int} vanishes when contracted by $p-p^\prime$.
Hence, it follows that
\be
\label{eq:continuity-eq}
\braket{N|\partial_\mu {\cal H}^{\mu\nu}|N^\star}=0 \ ,
\ee
which can be used to relate the matrix elements of the components of the tensor current.
Then 
the decay ratios for the transitions of a tensor boson
are explicitely computed and 
read
\begin{align}
R_\text{Be}=&\frac{k m_X^2}{18\pi}\Bigg|
\sqrt{\frac{4\pi}{3}} [(-\alpha_1+\beta_1 \xi)M1^\gamma_{I=1}\left(C_p-C_n\right) +\beta_1 M1^\gamma_{I=0}\left(C_p+C_n\right)] \nn \\
-&\frac{1}{2}
\left(5C_p-4C_n\right)\braket{\ce{^{8}Be}||\hat{\sigma}^{(p)}||\ce{^{8}Be}(18.15)} \\
+&\frac{1}{2}
\left(4C_p-5C_n\right)\braket{\ce{^{8}Be}||\hat{\sigma}^{(n)}||\ce{^{8}Be}(18.15)}
\Bigg|^2 \frac{\text{BR}(X\to e^+e^-)}{\Gamma({^8\text{Be}}(18.15)\to{^8\text{Be}}+\gamma)} \nn \ , \\
R_\text{He}=&\frac{m_N^2}{\alpha^2}\frac{5}{4}\frac{km_X^4}{\omega^5}\left( C_p+C_n \right)^2 \left|1-\left(3-2\frac{k^2}{m_X^2}\right)r_\text{He}\right|^2 \text{BR}(X\to e^+e^-) \ , \\
R_\text{C}=&\frac{m_N^2}{12\pi\alpha}\frac{m_X^4}{k\omega^3}\left[1+6r_\text{C}^2\right]\left(C_p-C_n\right)^2 \text{BR}(X\to e^+e^-) \ ,
\end{align}
where
\be
r_\text{C}=\frac{\braket{ \ce{^{12}C}||\sum_{s=1}^A(\vec{p}_s / m_N) \tau_z ||\ce{^{12}C}(17.63)}}{\braket{\ce{^{12}C}||\sum_{s=1}^A m_X \vec{r}_s\tau_z ||\ce{^{12}C}(17.63)}} \ ,
\ee
with $\tau_z=\text{diag}(1,-1)$ the third Pauli matrix in the isospin space.
For the case of the Helium and Carbon ratios, we have used the multipole expansion of the electromagnetic decay rates as calculated in Ref \cite{Barducci:2022lqd}, while the values of the parameters in the Beryllium ratio are reported in App.~\ref{App:chiSQ}.
%in Tab.\ \ref{tab:be_coeff}.
Finally, the branching ratio of the $X\to e^+ e^-$ decay channel for $S^\pi=2^+$ is given by
\be
\text{BR}(X\to e^+ e^-)=\frac{\Gamma(X\to e^+ e^-)}{\Gamma(X\to e^+ e^-)+\Gamma(X\to\gamma\gamma)}=\frac{C_e^2}{C_e^2+2C_\gamma^2} \ .
\ee

\subsection{Axial tensor boson $S^\pi=2^-$}

In this scenario,
parity conservation enforces $C_f=C_\gamma=0$.
From Eq.~\eqref{eq:nuc_int}, we derive the expressions of the nuclear tensor ${\cal H}^{\mu\nu}$ components at the leading order in the nonrelativistic limit, see App.~\ref{App:non_rel_spin2} for details.
By means of these components and the long-wavelength expansion as in Eq.~\eqref{eq:lwlexp}, we compute the decay ratios of the nuclear transitions for an axial tensor boson, in terms of reduced matrix elements of spherical operators.
We get
\begin{align}
R_\text{Be}=&\frac{m_N^2 k^3}{18\pi m_X^2}\Bigg| \tilde{C}_p\braket{\ce{^{8}Be}||\hat{\sigma}^{(p)}||\ce{^{8}Be}(18.15)}+\tilde{C}_n\braket{\ce{^{8}Be}||\hat{\sigma}^{(n)}||\ce{^{8}Be}(18.15)} \Bigg|^2 \nn \\
\times& \left(1+\frac{2}{3}\frac{\omega^2}{m_X^2}\right)\frac{\text{BR}(X\to e^+e^-)}{\Gamma({^8\text{Be}}(18.15)\to{^8\text{Be}}+\gamma)} \ , \\
R_\text{C}=&\frac{m_N^2}{32\pi\alpha}\frac{k^5}{m_X^2\omega^3}\left(\tilde{C}_p-\tilde{C}_n\right)^2\left|\tilde{r}_\text{C}\right|^2 \text{BR}(X\to e^+e^-) \ , \\
R_\text{He}=&\frac{80m_N^2}{\alpha^2}\frac{\sigma_- \Gamma_+}{\sigma_+ \Gamma_-}\left(\frac{k}{\omega}\right)^5\left( \tilde{C}_p+\tilde{C}_n \right)^2 |\tilde{r}_\text{He}|^2 \text{BR}(X\to e^+e^-) \ ,
\end{align}
where we introduced
\begin{align}
\tilde{r}_\text{He}=&\frac{\braket{ \ce{^{4}He}|| \sum_{s=1}^A \vec{\sigma}_s\cdot\vec{p}_s/m_N ||\ce{^{4}He}(21.01)}}{\braket{ \ce{^{4}He}|| \sum_{s=1}^A m_X^2 r_s^2||\ce{^{4}He}(20.21)}}  \ , \\
\tilde{r}_\text{C}=&\frac{\braket{\ce{^{12}C}||\sum_{s=1}^A (\vec{\sigma}_s\times\vec{r}_s) \tau_z ||\ce{^{12}C}(17.63)}}{\braket{\ce{^{12}C}||\sum_{s=1}^A\vec{r}_s \tau_z ||\ce{^{12}C}(17.63)}} \ .
\end{align}

In this case, the di-photon decay channel is suppressed with respect to the lepton one so one gets $\text{BR}(X\to e^+ e^-)\approx1$.

\section{Results}
\label{sec:res}

We now present our findings for the regions in the effective nucleon couplings parameter space for the various spin-parity assignments for the $X$ boson. 
To derive the range of values for the nucleons couplings, we construct a $\chi^2$ variable given by
\be
\chi^2
%=\sum_N\chi_N^2+\chi_\text{par}^2
=\sum_N\left(\frac{R_N-\mu_N^\text{exp}}{\sigma_N^\text{exp}}\right)^2
+\chi_\text{par}^2 \ ,
\ee
where $\mu_N^\text{exp}$ and $\sigma_N^\text{exp}$ are the experimental mean value and uncertainty recollected in Tab.\,\ref{tab:exp_value},
while $\chi_\text{par}^2$ is the contribution from the parameters involved in the theoretical expressions of $R_N$, see App.~\ref{App:chiSQ} for a detailed description.
The $\chi^2$ is then profiled over all the parameters, \emph{i.e.} at fixed values of the couplings we substitute the values of the parameters that minimize the chi-squared.
%Regarding the $r_{\text{He},\text{C}}$ and $\tilde{r}_{\text{He},\text{C}}$ parameters, which we do not include in $\chi_\text{par}^2$ because the theoretical uncertainty is not known, we restrict them to vary at most by a factor of 5 from its order of magnitude estimate, given in Tab.\,\ref{tab:estimate_par}.
We set $m_X=16.85$ MeV in all the calculations.

We emphasize that our analysis is based on several assumptions, such as the narrow-width approximation for the nuclear production of the excited states. Other possible contributions, such as direct capture processes, could alter the conclusion of our review.

\subsection{Axial tensor boson $S^\pi=2^-$}

\begin{figure}[t!]
\begin{center}
\includegraphics[scale=0.4]{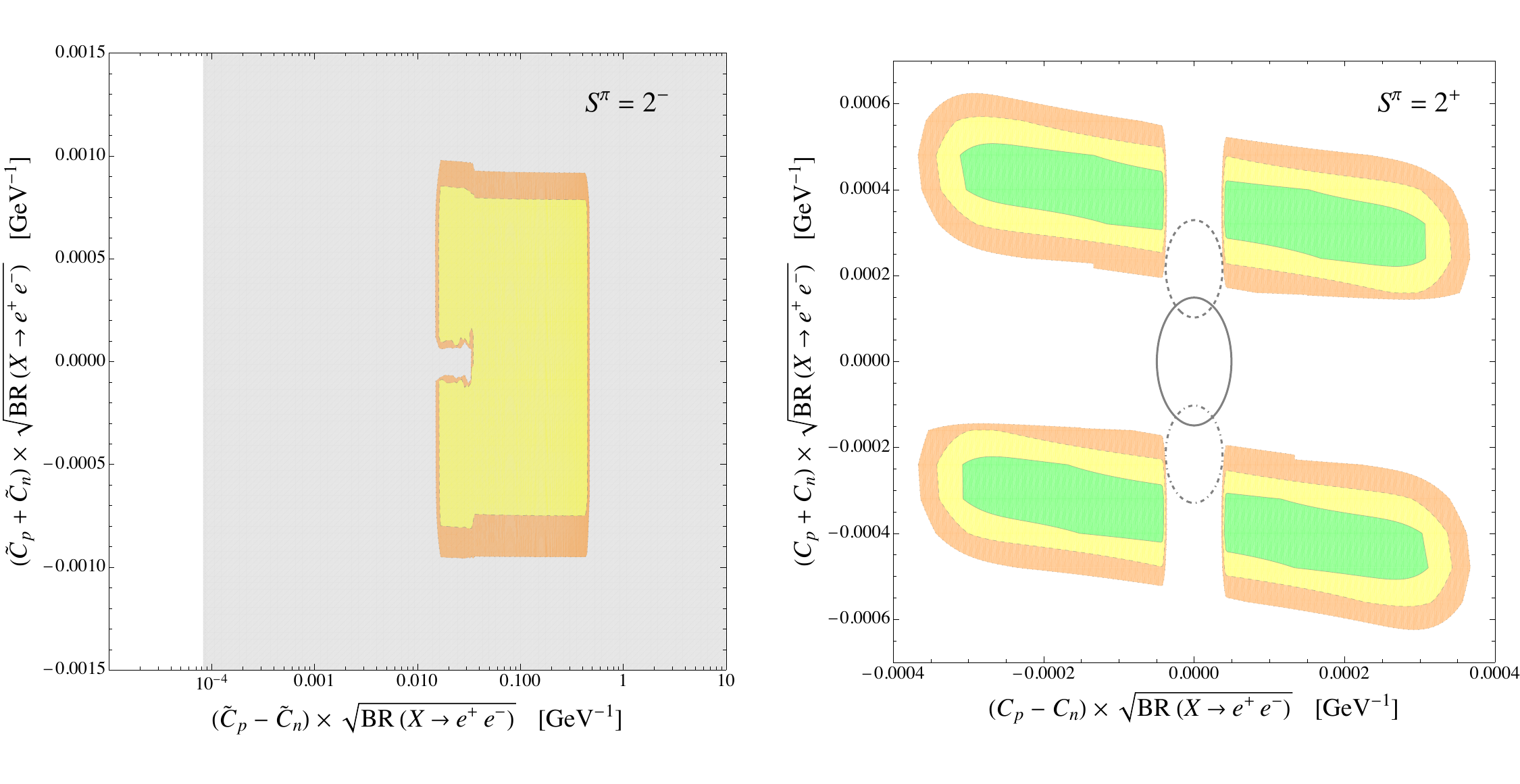}
\caption{\emph{Left panel:} Green, yellow, orange areas correspond to the $1\sigma,2\sigma,3\sigma$ compatibility regions, defined by the requirement $\chi_\text{profiled}^2<2.28,5.99,11.62$, for an axial tensor boson. The gray region is excluded by SINDRUM search. \emph{Right panel:} Green, yellow, orange areas correspond to the $1\sigma,2\sigma,3\sigma$ compatibility regions, defined by the requirement $\chi_\text{profiled}^2<2.28,5.99,11.62$, for a tensor boson. The regions outside the solid, dashed and dot-dashed gray lines are excluded by the SINDRUM search at 90\% CL respectively for $C_e=0$, $C_e=-0.001\text{ GeV}^{-1}$ and $C_e=0.001\text{ GeV}^{-1}$.}
\label{fig:spin2}
\end{center}
\end{figure}

We summarize the results for the tensor scenario in the left panel of Fig.~\ref{fig:spin2}, where we set, with no loss of generality,  $\tilde{C}_p-\tilde{C}_n>0$.
As one can see, the Atomki anomalies can be explained at best at $2\sigma$ and require an isoscalar coupling suppressed with respect to the isovector combination.
In gray, we also show the region excluded by the experimental search from the SINDRUM collaboration of the charged pion decay $\pi^+\to e^+ \nu_e X$ with $X\to e^+e^-$, whose bound reads \cite{SINDRUM:1986klz,Hostert:2023tkg}
\be
\text{BR}(\pi^+\to e^+ \nu_e X) \ \text{BR}(X\to e^+ e^-)<6.0\times10^{-10} \quad \text{at 90\% CL} \ ,
\ee
for a $X$ boson of mass around 17 MeV. The theoretical prediction is
\be
\text{BR}(\pi^+\to e^+\nu_e X)=
\frac{m_\pi^{12} \left(54 \eta_2^2 (\tilde{C}_p-\tilde{C}_n)^2+5 \tilde{C}_e^2\right)}{2^8 \ 3^2 \ 5^2 \pi ^2 m_\mu^2 m_X^4
   \left(m_\pi^2-m_\mu^2\right)^2}
\ee
where $\eta_2$ is a EFT coefficient that we expect of order of ${\cal O}(1)$, see details in App.~\ref{App:sindrum}, and from which we extract
\be
|\tilde{C}_p-\tilde{C}_n|\times \sqrt{\text{BR}(X\to e^+ e^-)}\lesssim {\cal O}(8.8\times10^{-5}) \ .
\ee
Hence, we concluded that the SINDRUM constraint completely excludes the axial tensor solution by orders of magnitude.

\subsection{Tensor boson $S^\pi=2^+$}

We summarize the results for the pure tensor scenario in the right panel of Fig.~\ref{fig:spin2}. There are regions of the parameter space that could accommodate all the nuclear anomalies within $1\sigma$. 
For the $S^\pi = 2^+$ case, the theoretical prediction of the charged pion decay $\pi^+\to e^+ \nu X$ is
\be
\text{BR}(\pi^+\to e^+\nu_e X)=\frac{m_\pi^{12} \left(90 \left(\eta_2^2 (C_p-C_n)^2+ \eta_3^2 (C_p+C_n)^2\right)+3 C_e^2+10 C_e \eta_3 (C_p+C_n)\right)}{2^8 \ 3^3 \ 5 \pi ^2 m_\mu^2 m_X^4 \left(m_\pi^2-m_\mu^2\right)^2} \ ,
\ee
where, again, $\eta_3$ is a EFT coefficient we expect of order of ${\cal O}(1)$, see details in App.~\ref{App:sindrum}.
In the regions of the $(C_p-C_n,C_p+C_n)$ plane that satisfy the Atomki anomalies, the SINDRUM bound leads to an upper limit on the electron coupling that reads $|C_e|\times \sqrt{\text{BR}(X\to e^+ e^-)}\lesssim10^{-3}$ GeV$^{-1}$.
We plot the contours of the regions excluded by the SINDRUM constraint in gray, both for $C_e=0$ (solid line) and for the maximum allowed absolute value of $C_e$ (dashed and dot-dashed line).
In this case, the search from NA48 is not constraing as $\pi^0\to\gamma X$ is forbidden by $C$-symmetry if one assigns $C(X)=+1$ from Eq.~\eqref{eq:nuc_int}.
In conclusion, we observe that the SINDRUM disfavours this scenario, up to our ignorance about the $\chi$PT coefficients.

\subsection{Scalar boson $S^\pi=0^+$}

\begin{figure}[t!]
\begin{center}
\includegraphics[scale=0.4]{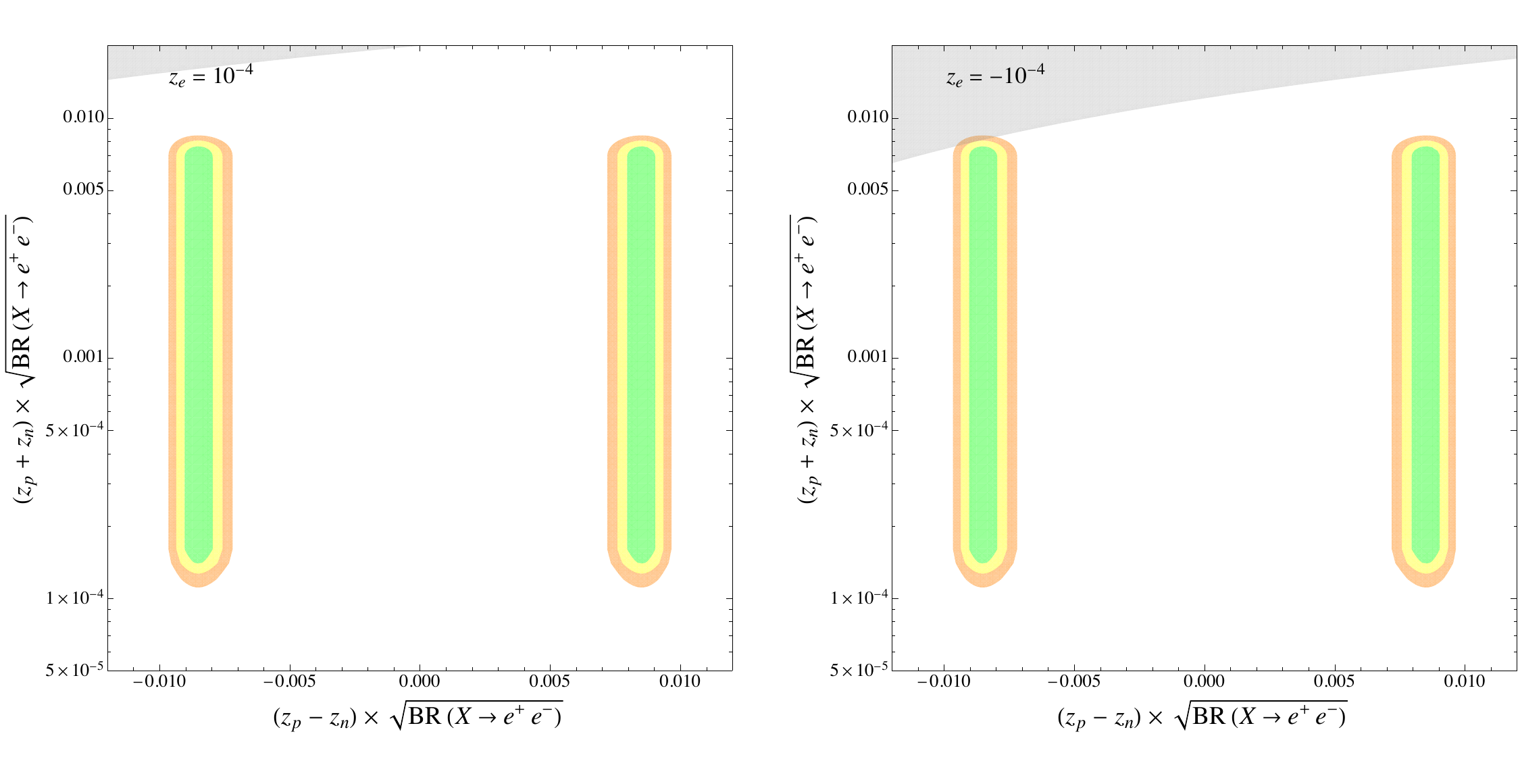}
\caption{Green, yellow, orange areas correspond to the $1\sigma,2\sigma,3\sigma$ compatibility regions, defined by the requirement $\chi_\text{profiled}^2<2.28,5.99,11.62$, for a scalar boson. The gray region is excluded by SINDRUM search fixing $z_e=10^{-4}$ (left panel) and $z_e=-10^{-4}$ (right panel).}
\label{fig:spin0p}
\end{center}
\end{figure}

In this case, we completely rely on the MEG-II result and assume that no anomalous signal is present in the Beryllium transition. Consequently, we completely discard its contribution to the $\chi^2$, as the transition is forbidden for a scalar boson due to parity conservation.
We summarize the results for the pure scalar scenario in Fig.~\ref{fig:spin0p}, where we set, with no loss of generality,  $z_p+z_n>0$.
Regarding the electron coupling, by asking that the BSM contribution to the electron's $g-2$ from the $X$ boson 
given by~\cite{Jegerlehner:2009ry}
\be
\delta a_e^{\rm BSM} = \frac{z_e^2}{4\pi^2}\frac{m_e^2}{m_X^2} \frac{1}{2}
\int_0^1 dz\; \frac{m_X^2 z^2(2-z)}{m_X^2 (1-z) + m_e^2 z^2}\approx \frac{z_e^2}{4\pi^2}\frac{m_e^2}{m_X^2}\left[\ln\frac{m_X}{m_e}-\frac{7}{12}\right] \ ,
\ee
does not overshoot the discrepancy between the central values of the SM prediction and the experimental measurement~\cite{Hanneke:2008tm} one obtains two different constraints, depending on the choice of the SM prediction
\begin{align}
\delta a_e^{\rm BSM}(\text{Rb}) &= (0.48\pm0.36) \times 10^{-12}  \ \ \text{\cite{Morel:2020dww}} \ , \\
\delta a_e^{\rm BSM}(\text{Cs}) &= (-0.88\pm0.30) \times 10^{-12} \ \ \text{\cite{Parker:2018vye}}  \ .
\end{align}
Conservatively, we set an upper limit by requiring
\be
|\delta a_e^{\rm BSM}|\leq\frac{\delta a_e^{\rm BSM}(\text{Rb}) - \delta a_e^{\rm BSM}(\text{Cs})}{2}=6.8\times10^{-13} \ \Longrightarrow \ |z_e|\lesssim10^{-4} \ .
\ee
A lower limit can be instead set by the requirement of a prompt decay within the Atomki detector, see App.~\ref{App:bound0} for details, which imposes
\be
|z_e|\gtrsim 1.4 \times10^{-5} \ .
\ee
We then fix the electron coupling to the maximal allowed value with a positive or negative sign and plot in gray the region excluded by the SINDRUM constraint, evaluated in App.~\ref{App:sindrum}, in Fig.~\ref{fig:spin0p}.
In constrast, the NA48 search does not constrain this scenario as $\pi^0\to\gamma X$ is forbidden by parity conservation for a scalar boson.
We then conclude the scalar boson could accomodate the Helium and Carbon anomalies even at the $1\sigma$ level while being compatible with all the experimental searches.
Finally, we comment on the result of the JINR collaboration. As previously mentioned, we reasonably expect the photon coupling of Eq.~\eqref{eq:nuc_int_0_g} to be suppressed, such that $\text{BR}(X\to \gamma\gamma)\ll1$, implying that a CP even scalar would unlikely produce a significant signal in JINR experiment.

\section{Conclusion}
\label{sec:conc}

The MEG-II experiment recently released the results of a measurement of the nuclear transition \( {^8}\text{Be}^\star \to {^8}\text{Be} + e^+e^- \)\,\cite{MEGII:2024urz}, showing no evidence of an anomalous signal. This contrasts with analogous measurements by the Atomki collaboration, which suggested the possible existence of a new boson \( X \) with a mass of approximately \( 17\,\text{MeV} \)\,\cite{Krasznahorkay:2015iga,Krasznahorkay:2018snd}. Meanwhile, the Atomki anomalies observed in \( ^4\text{He} \)\,\cite{Krasznahorkay:2019lyl,Krasznahorkay:2021joi} and \( ^{12}\text{C} \)\,\cite{Krasznahorkay:2022pxs} transitions remain to be independently verified. 

In light of this recent result, we have revisited the theoretical interpretations of the Atomki anomalies in terms of the proposed boson \( X \). We find that the measurements from the two experiments remain compatible within \( 2\sigma \), indicating that further data from MEG-II, expected in the near future, will be crucial to definitively rule out the Atomki Beryllium results. This compatibility also allows us to combine the two measurements, leading to a best-fit value for the Beryllium decay ratio, as defined in Sec.\,\ref{sec:review}, of \( R_\text{Be} = (5.5 \pm 1.0) \times 10^{-6} \). 

Building on this, we have extended the analysis in\,\cite{Barducci:2022lqd} by broadening the possible spin assignments of the \( X \) boson to include a spin-2 state with both positive and negative parity. However, we find that this explanation is strongly disfavored by experimental constraints from SINDRUM searches. Assuming future MEG-II results completely rule out the Beryllium anomaly, we have reconsidered the possibility of a pure CP-even scalar \( X \) state, a hypothesis previously dismissed due to its incompatibility with the Beryllium signal. Our results indicate that a pure CP-even scalar \( X \) can account for the Atomki anomalies in \( ^4\text{He} \) and \( ^{12}\text{C} \), while remaining consistent with other experimental constraints.

Further data from MEG-II, along with complementary experiments targeting \( ^4\text{He} \) and \( ^{12}\text{C} \) transitions, will be essential to confirm or refute this interpretation. 

\section*{Acknowledgements}

The work of MN and CT is supported by the Italian Ministry of University and Research (MUR) via the PRIN 2022 project n.~2022K4B58X -- AxionOrigins.
The work of CT has also received funding from the French ANR, under contracts ANR-19-CE31-0016 (`GammaRare') and ANR-23-CE31-0018 (`InvISYble'), that he gratefully acknowledges.
The authors thank A. D. Polosa for triggering this work and William Natale, Marco Mancini and Gianluca Cavoto for useful discussions.

%%%%%%%%%%%%%%%%%
%%%	Appendix	   %%%		
%%%%%%%%%%%%%%%%%

\appendix

\section{Single-nucleon operators of the tensor current}
\label{App:non_rel_spin2}

From Eq.~\eqref{eq:nuc_int}, we derive the leading terms of the nuclear tensor ${\cal H}^{\mu\nu}$ in the nonrelativistic limit. We label the tensor components as
\be
\mathcal{H}^{\mu\nu}=\begin{pmatrix}
\mathcal{E} & \vec{\mathcal{P}}^T \\
\vec{\mathcal{P}} &
\hat{\mathcal{W}}
\end{pmatrix} \ .
\ee

\subsubsection*{Axial tensor case $S^\pi = 2^-$}

For the case of a spin-2 with negative parity, the relevant single-nucleon operators are
\begin{align}
{\cal E}(\vec{r})=& \frac{1}{2}\sum_{s=1}^{A} \tilde{C}_s \left[\vec{p}_s \delta_{\vec{r},\vec{r}_s}+\delta_{\vec{r},\vec{r}_s}\vec{p}_s\right]\cdot\vec{\sigma}_s \ , \\
\vec{\cal P}(\vec{r})=&\frac{1}{2} \sum_{s=1}^{A} m_s\tilde{C}_s\vec{\sigma}_s \delta_{\vec{r},\vec{r}_s} \ , \\
%+& \sum_{s=1}^{A} \frac{\tilde{C}_s}{8m_s} \left[\vec{p}_s (\vec{p}_s\cdot\vec{\sigma}_s) \delta_{\vec{r},\vec{r}_s} + \vec{p}_s\delta_{\vec{r},\vec{r}_s}(\vec{p}_s\cdot\vec{\sigma}_s)\right] \nn \\
%+&\sum_{s=1}^{A} \frac{\tilde{C}_s}{8m_s}\left[(\vec{p}_s\cdot\vec{\sigma}_s)\delta_{\vec{r},\vec{r}_s}\vec{p}_s + \delta_{\vec{r},\vec{r}_s} \vec{p}_s(\vec{p}_s\cdot\vec{\sigma}_s) \right] \ , \\
\hat{\cal W}^{ij}(\vec{r})=&\frac{1}{4}\sum_{s=1}^{A} \tilde{C}_s \left[p_s^i \delta_{\vec{r},\vec{r}_s}+\delta_{\vec{r},\vec{r}_s}p_s^i\right]\sigma_s^j \nonumber \\
+&\frac{1}{4}\sum_{s=1}^{A} \tilde{C}_s \left[p_s^j \delta_{\vec{r},\vec{r}_s}+\delta_{\vec{r},\vec{r}_s}p_s^j\right]\sigma_s^i \ .
\end{align}

\subsubsection*{Tensor case $S^\pi = 2^+$}

For the case of a spin-2 with positive parity, the relevant single-nucleon operators are
\begin{align}
{\cal E}(\vec{r})=& \sum_{s=1}^{A} m_s C_s \delta_{\vec{r},\vec{r}_s} \nonumber \\
+&\sum_{s=1}^{A} \sum_i \frac{C_s}{8m_s}\left[p_s^i p_s^i \delta_{\vec{r},\vec{r}_s} + 2 p_s^i \delta_{\vec{r},\vec{r}_s}p_s^i + \delta_{\vec{r},\vec{r}_s} p_s^i p_s^i \right]\ , \\
\vec{\cal P}(\vec{r})=& \frac{1}{2}\sum_{s=1}^{A} C_s \left[\vec{p}_s \delta_{\vec{r},\vec{r}_s}+\delta_{\vec{r},\vec{r}_s}\vec{p}_s\right] +\vec{\nabla}\times\left(\frac{1}{4}\sum_{s=1}^{A} C_s \vec{\sigma}_s \delta_{\vec{r},\vec{r}_s}\right) \ , \\
\hat{\cal W}^{ij}(\vec{r})=&\sum_{s=1}^{A} \frac{C_s}{4m_s}\left[p_s^i p_s^j \delta_{\vec{r},\vec{r}_s} + p_s^i \delta_{\vec{r},\vec{r}_s}p_s^j + p_s^j\delta_{\vec{r},\vec{r}_s}p_s^i + \delta_{\vec{r},\vec{r}_s} p_s^i p_s^j \right] \nonumber \\
-&\sum_{s=1}^{A} \frac{C_s}{4m_s}(\vec{\sigma}_s\times\vec{\nabla})^j \left[p_s^i \delta_{\vec{r},\vec{r}_s} + \delta_{\vec{r},\vec{r}_s}p_s^i\right] \nonumber \\
-&\sum_{s=1}^{A} \frac{C_s}{4m_s}(\vec{\sigma}_s\times\vec{\nabla})^i \left[p_s^j \delta_{\vec{r},\vec{r}_s} + \delta_{\vec{r},\vec{r}_s}p_s^j\right] \ ,
\end{align}
where the next-to-leading terms of $\cal E$ are needed for the same reason exposed in the scalar case (see Sec.~\ref{sec_X17_0}).

\section{Details on the $\chi^2$ analysis}
\label{App:chiSQ}

The term $\chi^2_\text{par}$ is the sum of the constraints on the nuisance parameters involved in the theoretical expressions of the anomalous decay rates. Explicitly, it reads
\be
\chi^2_\text{par} = \sum_{\{p_i\}}\left(\frac{p_i-\mu_{p_i}}{\sigma_{p_i}}\right)^2 \ ,
\ee
where the sum is performed over the set of parameters
\be
\begin{split}
\{p_i\}=\Big\{&\alpha_1, \ \beta_1, \ \xi, \ M1^\gamma_{I=0}, \  M1^\gamma_{I=1}, \ \Gamma({^8\text{Be}}(18.15)\to{^8\text{Be}}+\gamma),\\
&\braket{\ce{^{8}Be}||\hat{\sigma}^{(p)}||\ce{^{8}Be}(18.15)}, \ \braket{\ce{^{8}Be}||\hat{\sigma}^{(n)}||\ce{^{8}Be}(18.15)}\Big\} \ .
\end{split}
\ee
Their mean values $\mu_{p_i}$ and uncertainties $\sigma_{p_i}$ are shown in Tab.~\ref{tab:be_coeff}.
\begin{table}[t!]
\begin{center}
\begin{tabular}{c|c|c|c|c}
$\alpha_1$ & $\beta_1$ & $\xi$ & $M1^\gamma_{I=0}$ & $M1^\gamma_{I=1}$ \\
\midrule
\midrule
$0.21(3)$ & $0.98(1)$ & $0.549$ & $ 0.014(1)$ & $0.767(9)$  \\
\end{tabular}
\end{center}
\vspace{-4mm}
\begin{center}
\begin{tabular}{c|c|c}
$\Gamma({^8\text{Be}}(18.15)\to{^8\text{Be}}+\gamma)$ & $\braket{\ce{^{8}Be}||\hat{\sigma}^{(p)}||\ce{^{8}Be}(18.15)}$ & $\braket{\ce{^{8}Be}||\hat{\sigma}^{(n)}||\ce{^{8}Be}(18.15)}$ \\
\midrule
\midrule
$1.9(4)$ keV & $-0.047(29)$ & $-0.132(33)$  \\
\end{tabular}
\end{center}
\caption{Values of the coefficients needed for the Beryllium ratios taken from Ref.~\cite{Barducci:2022lqd}.}
\label{tab:be_coeff}
\end{table}
Regarding the $r_{\text{He},\text{C}}$ and $\tilde{r}_{\text{He},\text{C}}$ parameters, we do not include them in $\chi_\text{par}^2$ because no theoretical calculation or experimental measurement of them is available in literature.
Instead, we decide to restrict them to vary at most by a factor of 5 from their order of magnitude estimates, given in Tab.\,\ref{tab:estimate_par}, while minimising the $\chi^2$ over the parameters at fixed coupling values.
\begin{table}[t!]
\begin{center}
\begin{tabular}{c|c|c|c}
$r_\text{He}$ & $\tilde{r}_\text{He}$ & $r_\text{C}$ & $\tilde{r}_\text{C}$ \\
\midrule
\midrule
$\sim4.6$ & $\sim7.7$ & $\sim5.5$ & $\sim1$  \\
\end{tabular}
\end{center}
\caption{Estimates of the unknown parameters by considering $\vec{p}_s\sim \braket{p_N}$ and $\vec{r}_s\sim \text{1.3 fm $A^{1/3}$}$.}
\label{tab:estimate_par}
\end{table}

\section{Charged pion decay $\pi^+\to e^+ \nu_e X$}
\label{App:sindrum}

\begin{figure}[t!]
\begin{center}
\includegraphics[scale=0.65]{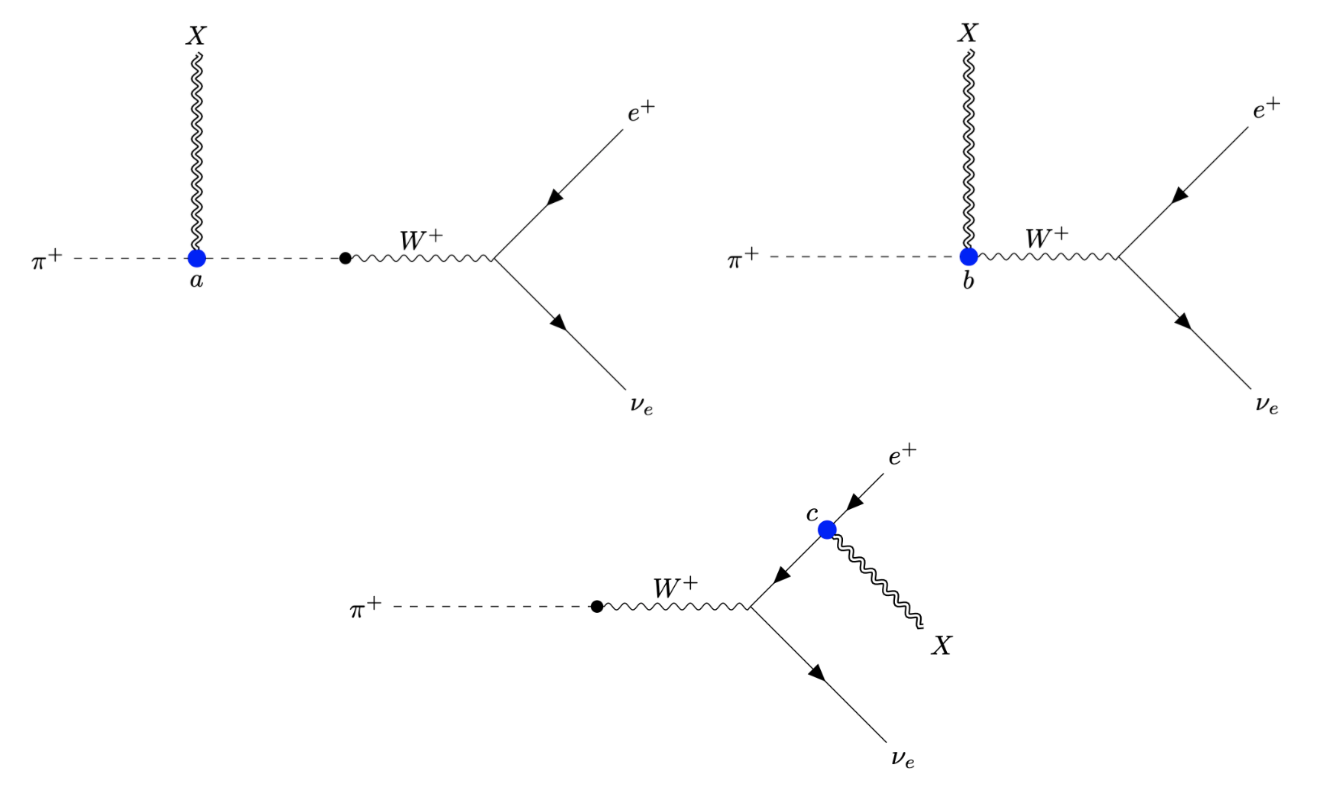}
\caption{Diagrams of internal bremsstrahlung of the charged decay pion.}
\label{fig:cpd}
\end{center}
\end{figure}

In this Appendix we derive the theoretical prediction of the charged pion decay
\be
\pi^+ (q)\to e^+(q_1)\, \nu_e(q_2)\, X(q_3)
\ee
for a boson with $S^\pi=0^+,2^+,2^-$.
The $X$ boson is produced by internal bremsstrahlung, see Fig.~\ref{fig:cpd}, and it could be emitted by the pion, the leptons and in the $W-\text{pion}$ vertex\footnote{It could be emitted also from the $W$ line but it would be doubly suppressed by the Fermi constant so one can safely discard it.}.
The interactions of the $X$ boson to the charged leptons is assumed according to Eqs.~\eqref{eq:nuc_int_0}--\eqref{eq:nuc_int} while we neglect a coupling to neutrinos for simplicity. Finally the weak lepton Lagrangian is given by
\be
\mathcal{L}_{\text{weak-lepton}}=\frac{g}{\sqrt{2}}\left( \overline{\nu}_{eL}\gamma^\mu e_L W_\mu^+ +  \overline{e}_L\gamma^\mu \nu_{eL} W_\mu^-\right) \ ,
\ee
while, instead, the interaction to the pion is parametrized within the framework of chiral pertubation theory ($\chi$PT) starting from the $X$ couplings to up and down quarks. For simplicity, we will not consider couplings to other quarks.

\subsection{Chiral pertubation theory}

The QCD Lagrangian in the massless limit
\begin{equation}
\mathcal{L}_{\text{QCD}}^0 = \overline{q}_L i\slashed{\partial} q_L + \overline{q}_R i\slashed{\partial} q_R -\frac{1}{4} G_{\mu\nu}^a G^{\mu\nu, a} \quad \text{with $q=\begin{pmatrix} u \\ d \end{pmatrix}$,}
\end{equation}
shows a global symmetry $SU(2)_L \times SU(2)_R \times U(1)_V$ under which the quark fields transform as
\be
q_{L} \to g_{L}\, q_{L} \ , \quad q_{R} \to g_{R}\, q_{R} \ ,
\ee
with $g_{L}\in SU(2)_{L}$ and $g_{R}\in SU(2)_{R}$. Furthermore, it is invariant under discrete parity symmetry.
At low energy, the quark density acquires a non vanishing value $\braket{\overline{q}q}\neq0$ and breaks spontaneausly the global symmetry as
\begin{equation}
SU(2)_L \times SU(2)_R \times U(1)_V \to SU(2)_V \times U(1)_V \ ,
\end{equation}
where the corrisponding Goldstone bosons are identified as the pions. In $\chi$PT the pions fields are embedeed in the matrix
\be
U =
\text{exp}\left\{
\frac{i}{f_\pi}
\begin{pmatrix}
\pi^0 & \sqrt{2}\pi^+ \\
\sqrt{2}\pi^- & -\pi^0
\end{pmatrix}
\right\} \ ,
\ee
transforming under the global symmetry as
\be
U\to g_R U g_L^\dagger \ ,
\ee
where $f_\pi=92.4\pm0.3$ MeV is the pion decay constant.
Symmetry breaking terms, as the quark masses, and gauge interactions are then introduced as external currents in QCD Lagrangian as
\be
\mathcal{L}_{\text{QCD}} = \mathcal{L}_{\text{QCD}}^0
+ \overline{q}_L \gamma^\mu l_\mu q_L + \overline{q}_R \gamma^\mu r_\mu q_R - \overline{q}_L(s-ip)q_R -\overline{q}_R(s+ip)q_L \ ,
\ee
and promoting the global symmetry to be local.
The external currents trasform under local chiral rotations as
\begin{align}
&l_\mu \to g_L \,l_\mu\, g_L^\dagger + ig_L\, \partial_\mu g_L^\dagger \ , \\
&r_\mu \to g_R \,r_\mu\, g_R^\dagger + ig_R\, \partial_\mu g_R^\dagger \ , \\
&(s+ip) \to g_R(s+ip)g_L^\dagger \ .
\end{align}
At lowest order, the $\chi$PT Lagrangian is given by \cite{Pich:1998xt}
\be
\mathcal{L}_{\text{$\chi$PT}}=
\frac{f_\pi^2}{4}\text{Tr}\left[\left(D^\mu U\right)^\dagger D_\mu U \right]+
\frac{f_\pi^2}{4}\text{Tr}\left[U^\dagger \chi + \chi^\dagger U \right] ,
\ee
where we introduced the notation
\begin{align}
D_\mu U &= \partial_\mu U + i U l_\mu - i r_\mu U \ , \\
\chi&=2B_0 (s+ip) \ ,
\end{align}
with the parameter $B_0$ related to quark condensate. Finally, we specify the external currents to be
\begin{align}
&s+ip=
\begin{pmatrix}
m_u & 0 \\
0 & m_d
\end{pmatrix}
-
X\begin{pmatrix}
z_u & 0 \\
0 & z_d
\end{pmatrix} \ , \nn \\
& l_\mu = \frac{g}{\sqrt{2}}
\begin{pmatrix}
0 & V_{ud} W_{\mu}^{+} \\
V_{ud}^* W_{\mu}^{-} & 0
\end{pmatrix} \ , \\
& r_\mu = 0 \ , \nn
\end{align}
with $X$ the $S^\pi=0^+$ boson field and
where we neglect terms which are irrelevant for the present discussion.
Following the same logic, we consider spin-2 interactions to quark through external tensor (symmetric) currents as
\be
\begin{split}
\Delta\mathcal{L}_{\text{QCD}}^\text{spin-2}=&\frac{i}{2}\overline{q}_L \gamma_\nu  \chi_L^{\mu\nu} (D_\mu q_L) - \frac{i}{2}(D_\mu\overline{q}_L) \gamma_\nu  \chi_L^{\mu\nu} q_L \\
+&\frac{i}{2}\overline{q}_R \gamma_\nu  \chi_R^{\mu\nu} (D_\mu q_R) - \frac{i}{2}(D_\mu\overline{q}_R) \gamma_\nu  \chi_R^{\mu\nu} q_R
\end{split}
\ee
with $D_\mu q_L=\partial_\mu q_L-il_\mu q_L$, $D_\mu q_R=\partial_\mu q_R-ir_\mu q_R$ while $\chi_R^{\mu\nu}$ and $\chi_L^{\mu\nu}$ transform as
\be
\chi_L^{\mu\nu} \to g_L \chi_L^{\mu\nu} g_L^\dagger \qquad \text{and} \qquad \chi_R^{\mu\nu} \to g_R \chi_R^{\mu\nu} g_R^\dagger \ .
\ee
At the lowest order in $\chi$PT, we then get
\be
\begin{split}
\Delta\mathcal{L}_{\text{$\chi$PT}}^\text{spin-2}=\frac{f_\pi^2}{4}\Biggr\{
&(\eta_1+i\eta_2) \text{Tr}\left[(D_\mu U)U^\dagger(D_\nu\chi_L^{\mu\nu})\right]
+(\eta_1+i\eta_2) \text{Tr}\left[(D_\mu U)^\dagger U(D_\nu\chi_R^{\mu\nu})\right] \\
+&(\eta_1-i\eta_2) \text{Tr}\left[U(D_\mu U)^\dagger(D_\nu\chi_L^{\mu\nu})\right]
+(\eta_1-i\eta_2) \text{Tr}\left[U^\dagger (D_\mu U)(D_\nu\chi_R^{\mu\nu})\right] \\
+&\eta_3 \text{Tr}\left[(D_\nu U)(D_\mu U)^\dagger \chi_L^{\mu\nu}\right]
+\eta_3 \text{Tr}\left[(D_\nu U)^\dagger (D_\mu U)\chi_R^{\mu\nu}\right]
\Biggr\} \ ,
\end{split}
\ee
where $\eta_{1,2,3}$ are adimensional real EFT coefficients while $D_\alpha\chi_L^{\mu\nu}=\partial_\alpha\chi_L^{\mu\nu}+i\chi_L^{\mu\nu}l_\alpha-il_\alpha \chi_L^{\mu\nu}$ and $D_\alpha\chi_R^{\mu\nu}=\partial_\alpha\chi_R^{\mu\nu}+i\chi_R^{\mu\nu}r_\alpha-ir_\alpha \chi_R^{\mu\nu}$.
Note that $\Delta\mathcal{L}_{\text{$\chi$PT}}^\text{spin-2}$ has the same energy counting of $\mathcal{L}_{\text{$\chi$PT}}$ and we choose the same normalization factor $f_\pi^2/4$, thus we expect $\eta_i$ to be of ${\cal O}(1)$.
According to the definition in Eq.~\eqref{eq:nuc_int}, the tensor currents are then specified to be
\be
\chi_L^{\mu\nu}= \begin{pmatrix}
C_u -\tilde{C}_u & 0 \\
0 & C_d -\tilde{C}_d
\end{pmatrix}X^{\mu\nu}
\qquad \text{and} \qquad
\chi_R^{\mu\nu}= \begin{pmatrix}
C_u +\tilde{C}_u & 0 \\
0 & C_d +\tilde{C}_d
\end{pmatrix}X^{\mu\nu} \ .
\ee

\subsection{Internal bremsstrahlung for a spin-2 boson}

Expanding the pion matrix in the $\chi$PT Lagrangian, we obtain the whole set of terms we need, \emph{i.e}
\be
\begin{split}
\mathcal{L}_{\text{$\chi$PT}}+\Delta\mathcal{L}_{\text{$\chi$PT}}^\text{spin-2} \supset \ & (\partial_\mu \pi^+)(\partial^\mu \pi^-)-m_\pi^2 \pi^+ \pi^- \\
&+\eta_3(C_u+C_d)X^{\mu\nu}(\partial_\mu \pi^+)(\partial_\nu \pi^-) \\
&+ \frac{gf_\pi}{2}\eta_3(C_u+C_d)X^{\mu\nu}(V_{ud}W_\mu^+ \partial_\nu \pi^- + V_{ud}^*W_\mu^- \partial_\nu \pi^+) \\
&-i\frac{gf_\pi}{2}\eta_2(C_u-C_d)X^{\mu\nu}(V_{ud}W_\mu^+ \partial_\nu \pi^- - V_{ud}^*W_\mu^- \partial_\nu \pi^+) \\
&+i\frac{gf_\pi}{2}\eta_2(\tilde{C}_u-\tilde{C}_d)X^{\mu\nu}(V_{ud}W_\mu^+ \partial_\nu \pi^- - V_{ud}^*W_\mu^- \partial_\nu \pi^+) \\
&+\frac{gf_\pi}{2}(V_{ud}W_\mu^+ \partial_\nu \pi^- + V_{ud}^*W_\mu^- \partial_\nu \pi^+) + \dots \ ,
\end{split}
\ee
with $m_\pi^2=B_0(m_u+m_d)$.
In the limit of point-like $W$ propagator, \emph{i.e.}
\be
\frac{i}{p^2-m_W^2}\left(-g_{\mu\nu}+\frac{p_\mu p_\nu}{m_W^2}\right)\approx\frac{i}{m_W^2}g_{\mu\nu} \ ,
\ee
the invariant matrix elements of each possible insertion of the $X$ boson are respectvely given by
\begin{align}
i{\cal M}_a &=-2V_{ud}^* m_e G_F f_\pi (C_u+C_d)\eta_3 [\epsilon_{\alpha\beta}(q_3)]^* q^\alpha q^\beta \ \overline{u}(q_2) P_R v(q_1) \ \frac{1}{(q-q_3)^2-m_\pi^2} \ , \\
i{\cal M}_b &=-2V_{ud}^* G_F f_\pi [\epsilon_{\alpha\beta}(q_3)]^* q^\alpha \ \overline{u}(q_2)\gamma^\beta P_L v(q_1) \left[(C_u+C_d)\eta_3+i\eta_2(C_u+\tilde{C}_d-C_d-\tilde{C}_u)\right] \ , \\
i{\cal M}_c &=-2V_{ud}^* G_F f_\pi [\epsilon_{\alpha\beta}(q_3)]^* q_1^\beta  \ \overline{u}(q_2)\slashed{q}P_L\frac{1}{\slashed{q}_2+\slashed{q}-m_e}\gamma^\alpha(C_e + \tilde{C}_e \gamma_5) v(q_1) \ ,
\end{align}
where we introduced the Fermi constant
\be
\frac{G_F}{\sqrt{2}}=\frac{g^2}{8m_W^2} \ .
\ee
The decay width is
\be
\Gamma(\pi^+\to e^+\nu_e X) = \frac{1}{(2\pi)^3\,32m_{\pi}^3}\int dm_{12}^2 dm_{23}^2  \ \sum_{\text{pol}}|{\cal M}|^2 \ ,
\ee
with ${\cal M}={\cal M}_a+{\cal M}_b+{\cal M}_c$ and $m_{ij}^2=(q_i+q_j)^2$, to be compared to
\be
\Gamma_{\pi^+}\approx\Gamma(\pi^+\to\mu^+\nu_\mu) = \frac{G_F^2f_\pi^2m_\mu^2m_\pi}{4\pi}|V_{ud}|^2\left(1-\frac{m_\mu^2}{m_\pi^2}\right)^2 \ .
\ee
Performing the integral in the limit $m_e,m_X\ll m_\pi$, we get
\be
\text{BR}(\pi^+\to e^+\nu_e X)=\frac{m_\pi^{12} \left(10 \left(\eta_2^2 (C_d-C_u)^2+ \eta_3^2 (C_u+C_d)^2\right)+3 C_e^2-10 C_e \eta_3 (C_u+C_d)\right)}{2^8 \ 3^3 \ 5 \pi ^2 m_\mu^2 m_X^4 \left(m_\pi^2-m_\mu^2\right)^2}
\ee
for a tensor boson and
\be
\text{BR}(\pi^+\to e^+\nu_e X)=
\frac{m_\pi^{12} \left(10 \eta_2^2 (\tilde{C}_d-\tilde{C}_u)^2+3 \tilde{C}_e^2\right)}{2^8 \ 3^3 \ 5 \pi ^2 m_\mu^2 m_X^4
   \left(m_\pi^2-m_\mu^2\right)^2}
\ee
for an axial tensor boson.

\subsubsection{Spin-2 nucleon couplings in the static quark model}

We derived the branching ratios in terms of the couplings to quarks. To connect them to the couplings to nucleons, which are involved in the Atomki transitions, we rely on the static quark model.

The static quark model describes nucleons as quark states $\ket{q_1 \uparrow,q_2\uparrow,q_3\downarrow}$,
where the arrows indicate spin orientation, in all the flavor and spin combinations. Identical quarks are in a $J=1$ spin state, such that spin combination reads as
\be
\ket{1/2,1/2}=\sqrt{2/3}\ket{1,1;1/2,-1/2}-\sqrt{1/3}\ket{1,0;1/2,1/2} \ .
\ee
In Ref.~\cite{dispense:2018} it is shown how this model actually leads to a good estimate of the nucleon magnetic moments.
For a $S^\pi=2^+$ boson, one gets at low-energy the operator ${\cal E}\approx m_s C_s \mathbf{1}$ so
\be
\begin{split}
C_p = \frac{1}{m_N} \bra{p}{\cal E}\ket{p}=\frac{2}{3}\frac{2m_u^\text{eff}C_u+m_d^\text{eff}C_d}{m_N}+\frac{1}{3}\frac{2m_u^\text{eff}C_u+m_d^\text{eff}C_d}{m_N}=\frac{2}{3}C_u+\frac{1}{3}C_d \ , \\
C_n = \frac{1}{m_N} \bra{n}{\cal E}\ket{n}=\frac{2}{3}\frac{m_u^\text{eff}C_u+2m_d^\text{eff}C_d}{m_N}+\frac{1}{3}\frac{m_u^\text{eff}C_u+2m_d^\text{eff}C_d}{m_N}=\frac{1}{3}C_u+\frac{2}{3}C_d \ ,
\end{split}
\ee
where $m_u^\text{eff}=m_u^\text{eff}=m_N/3$ are the effective quark masses.
Instead for a $S^\pi=2^-$ boson, we consider at low-energy the operator $\vec{\cal P}\approx m_s \tilde{C}_s \vec{\sigma}_s$, so
\be
\begin{split}
\tilde{C}_p = \frac{1}{m_N} \bra{p}{\cal P}\ket{p}=\frac{2}{3}\frac{2m_u^\text{eff}\tilde{C}_u-m_d^\text{eff}\tilde{C}_d}{m_N}+\frac{1}{3}\frac{m_d^\text{eff}\tilde{C}_d}{m_N}=\frac{4}{9}\tilde{C}_u-\frac{1}{9}\tilde{C}_d \ , \\
\tilde{C}_n = \frac{1}{m_N} \bra{n}{\cal P}\ket{n}=\frac{2}{3}\frac{-m_u^\text{eff}\tilde{C}_u+2m_d^\text{eff}\tilde{C}_d}{m_N}+\frac{1}{3}\frac{m_u^\text{eff}\tilde{C}_u}{m_N}=-\frac{1}{9}\tilde{C}_u+\frac{4}{9}\tilde{C}_d \ .
\end{split}
\ee

\subsection{Internal bremsstrahlung for a scalar boson}

In the limit of point-like $W$ propagator, the invariant matrix elements are respectevely given by
\begin{align}
i{\cal M}_a &=2V_{ud}^*m_e G_F f_\pi m_\pi^2 \frac{z_u+z_d}{m_u+m_d} \ \overline{u}(q_2) P_R v(q_1) \ \frac{1}{(q-q_3)^2-m_\pi^2} \ , \\
i{\cal M}_b &=0 \ , \\
i{\cal M}_c &=2V_{ud}^* z_e G_F f_\pi   \ \overline{u}(q_2)\slashed{q}P_L\frac{1}{\slashed{q}_2+\slashed{q}-m_e}v(q_1) \ .
\end{align}
However, since the first term is helicity suppressed, the next-to-leading order (NLO) in $\chi$PT accidentally gives a contribution of the same size of the leading order, which we must then take into account.
The NLO order $\chi$PT Lagrangian is given by \cite{Pich:1998xt}
\be
\label{eq:NLOchi}
\begin{split}
\mathcal{L}_{\text{$\chi$PT}}^\text{NLO}=& L_1 \text{Tr}\left[D_\mu U^\dagger D^\mu U\right]^2
+ L_2\text{Tr}\left[D_\mu U^\dagger D_\nu U\right]\text{Tr}\left[D^\mu U^\dagger D^\nu U\right] \\
+&L_3 \text{Tr}\left[D_\mu U^\dagger D^\mu U D_\nu U^\dagger D^\nu U\right] + L_4 \text{Tr}\left[D_\mu U^\dagger D^\mu U\right]\text{Tr}\left[U^\dagger \chi + \chi^\dagger U\right] \\
+& L_5 \text{Tr}\left[D_\mu U^\dagger D^\mu U(U^\dagger \chi + \chi^\dagger U)\right] + L_6 \text{Tr}\left[U^\dagger \chi + \chi^\dagger U\right]^2 \\
+& L_7 \text{Tr}\left[U^\dagger \chi - \chi^\dagger U\right]^2
+ L_8 \text{Tr}\left[U^\dagger \chi U^\dagger \chi + \chi^\dagger U \chi^\dagger U\right] + \dots \ ,
\end{split}
\ee
where the dots refer to operators we are not interested in for the present discussion. At the NLO tree level, one has to consider also the 1-loop contribution from the leading order $\chi$PT Lagrangian whose divergences are absorbed in the NLO $\chi$PT coefficients. The renormalized values of $L_i^r$ evalueted at the $\rho$ mass are recollected in Ref.~\cite{Pich:1998xt} and are order of ${\cal O}(10^{-3})$ as expected from naive EFT estimate, see Ref.~\cite{Pich:1998xt} for a detailed discussion. The running of the couplings is given by
\be
L_i^r(\mu_2) = L_i^r(\mu_1)+\frac{\Gamma_i}{16\pi^2}\ln\frac{\mu_1}{\mu_2} \ ,
\ee
with $\Gamma_i$ given as in Ref.~\cite{Pich:1998xt}.
The NLO operators in Eq.~\eqref{eq:NLOchi} contribute to the kinetic and mass pion terms so we need to rescale the pion fields and mass as
\be
\pi^\pm\to(1-\delta_2)\pi^\pm \quad \text{and} \quad m_\pi^2 \to (1+2\delta_2-\delta_1) m_\pi^2  \ ,
\ee
which leads to
\be
\begin{split}
\mathcal{L}_{\text{$\chi$PT}}+\mathcal{L}_{\text{$\chi$PT}}^\text{NLO}\left. \right|_\text{rescaled} \supset \ & (\partial_\mu \pi^+)(\partial^\mu \pi^-)-m_\pi^2 \pi^+ \pi^- \\
&+(1+\delta_1) m_\pi^2\frac{z_u+z_d}{m_u+m_d}X\pi^+\pi^- \\
&-2\delta_2\frac{z_u+z_d}{m_u+m_d}X(\partial_\mu \pi^+)(\partial^\mu \pi^-) \\
&-\delta_2 gf_\pi\frac{z_u+z_d}{m_u+m_d}X(V_{ud}W_\mu^+ \partial^\mu \pi^- + V_{ud}^*W_\mu^- \partial^\mu \pi^+) \\
&+\frac{gf_\pi}{2}(1+\delta_2)(V_{ud}W_\mu^+ \partial^\mu \pi^- + V_{ud}^*W_\mu^- \partial^\mu \pi^+) + \dots \ ,
\end{split}
\ee
where we neglect terms order ${\cal O}(\delta_i \delta_j)$, with
\be
\delta_1 = 16\frac{m_\pi^2}{f_\pi^2}\big[2L_6(m_\pi)+L_8(m_\pi)\big] \quad \text{and} \quad \delta_2=4\frac{m_\pi^2}{f_\pi^2}\big[2L_4(m_\pi)+L_5(m_\pi)\big] \ .
\ee
The total decay width is then given by
\be
\begin{split}
\Gamma(\pi^+\to e^+\nu_e X)= \frac{G_F^2 f_\pi^2 m_\pi^3}{32(2\pi)^3}|V_{ud}|^2\left[(z_u+z_d)^2F_1+z_e(z_u+z_d)F_2+z_e^2F_3\right] \ ,
\end{split}
\ee
where $F_{i}=F_{i}(m_X^2/m_\pi^2,m_e^2/m_\pi^2)$ are form factors with complex and involved expressions. We numerically evaluated
\begin{align}
F_1&\approx0.024 \ , \nn \\
F_2&\approx-0.144 \ , \\
F_3&\approx0.676 \nn \ .
\end{align}

Finally, the dependence of the nucleon couplings to the quark ones is obtained as described in App. C of Ref.~\cite{Barducci:2022lqd}. The nucleon effective couplings read as
\begin{equation}
z_{N}=m_N\sum_{q=u,d}\frac{z_q}{m_q}f_{Tq}^{(N)} \ ,
\end{equation}
for $N=p,n$,  with \cite{Ellis:2000ds}
\be
\begin{split}
& f_{Tu}^{(p)}=0.020\pm0.004, \quad f_{Td}^{(p)}=0.026\pm0.005 \ , \\
& f_{Tu}^{(n)}=0.014\pm0.003,\quad f_{Td}^{(n)}=0.036\pm0.008 \ .
\end{split}
\ee

\section{Prompt decay in Atomki}
\label{App:bound0}

In this Appendix we evaluate the bound on the decay width of the $X$ boson coming from the requirement of a prompt decay in the Atomki experiment.
Indeed, the Atomki collaboration evaluate the mass of the $X$ boson by a fit of the angular correlation among the detected lepton pairs \cite{Krasznahorkay:2015iga, Barducci:2022lqd,Denton:2023gat}. 

\begin{figure}[t]
\centering
\includegraphics[width=0.5\textwidth]{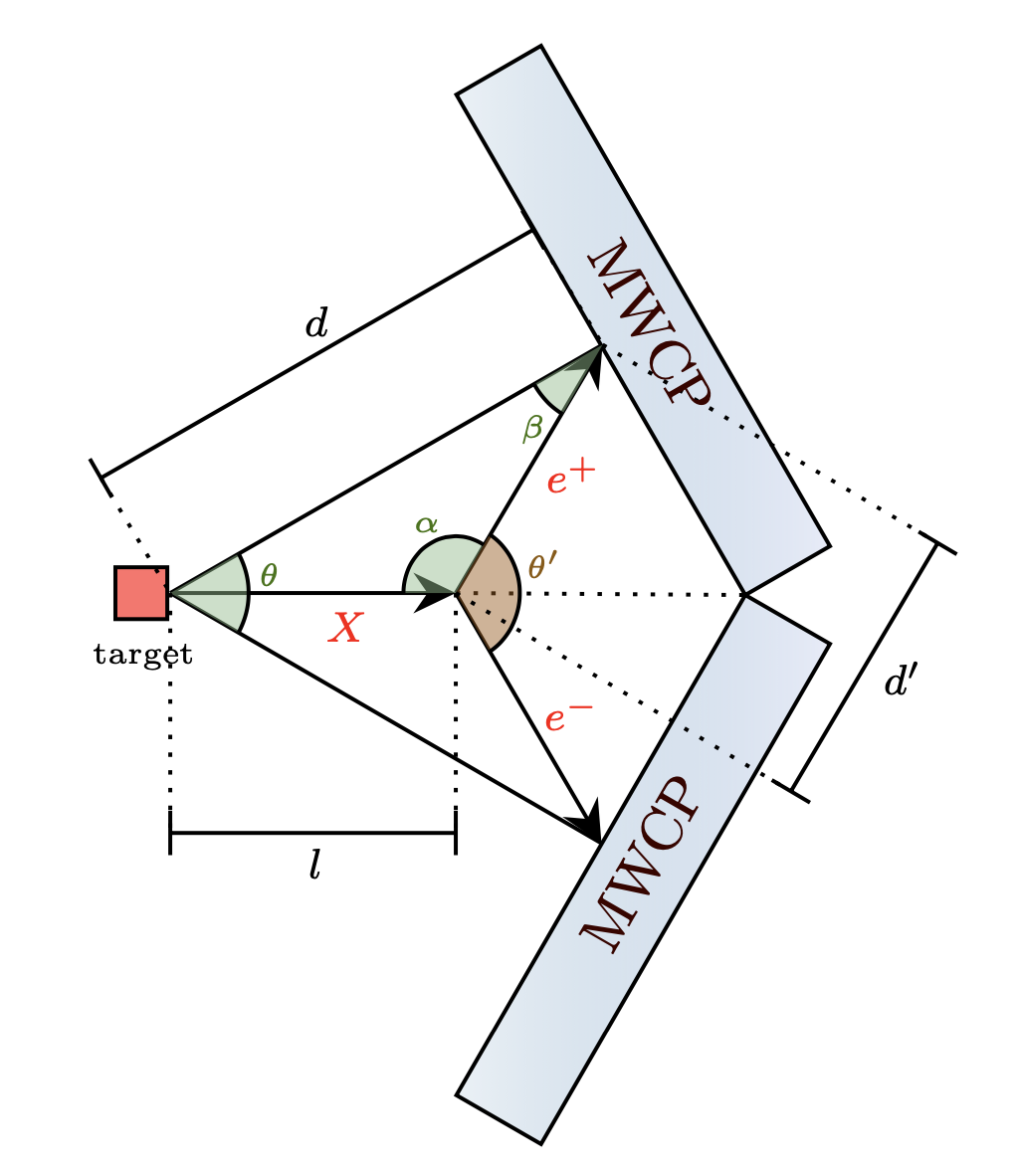}
\caption{\small Effect of a long decaying resonance on angle correlation. The figure replicate the Atomki setup as shown in Ref.~\cite{Gulyas:2015mia}.}
\label{Atomki distortion}
\end{figure}

However, once produced at the interaction point, the $X$ boson can actually travel inside the apparatus before decaying, which can potentially smear the invariant mass distribution leading to a disappearance of the anomalous peak itself. This smearing introduced by the decay length of the particle must be contained within the resolution bounds of the opening angle, otherwise the peak would disappear. The geometry of the apparatus is shown in Fig.~\ref{Atomki distortion}, where $\theta^\prime$ is the opening angle among the leptons while $\theta$ is the angle reconstracted from the position of the hits measured by the multiwire proportional counters (MWCP) located at distance $d=5$ cm from the target.

The distance $l$ the $X$ boson travels before it decays is obtained applying the law of sins
\be
\frac{\sin\beta}{l}=\frac{\sin\alpha}{d} \qquad \Longrightarrow \qquad l=\frac{\sin\beta}{\sin\alpha}d =\frac{\sin(\Delta\theta/2)}{\sin(\theta^\prime/2)}d 
\ee
and we get a constraint on this distance by requiring $\Delta \theta \equiv \theta' - \theta < \Delta \theta_{\text{exp}}=\ang{2}$ \cite{Gulyas:2015mia}, while varying over the values of $\theta^\prime$. The total decay width of the boson is subsequently obtained as $\Gamma=\gamma\beta/l$, with $\beta$ the boson velocity and $\gamma=(1-\beta^2)^{-1/2}$ its Lorentz factor.
We recollect the constraint thus derived for each Atomki experiment in Tab.\,\ref{table Atomki}. The strongest bound comes from the Helium experiment and reads
\be
\Gamma \geq 1.3 \times10^{-10} \text{ MeV} \ .
\ee

\begin{table}[ht]
\centering
\begin{tabular}{c|c|c|c|c|c}
\hline
\hline
    Experiment & $\omega$ [MeV] & $\beta$ & $\gamma$ & $l_\text{max}$ [cm] & $\Gamma_\text{min}$ [MeV] \\
    \hline
    \hline
    $^8$Be & 18.15 & 0.359 & 1.071 & 0.092 & $8.2\times10^{-11}$ \\
    \hline
    $^4$He & 20.49 & 0.569 & 1.216 & 0.104 & $1.3\times10^{-10}$ \\
    \hline
    $^{12}$C & 17.23 & 0.186 & 1.018 & 0.088 & $4.3\times10^{-11}$ \\
    \hline 
    \hline
\end{tabular}
\caption{Values of the boson energy $\omega$, velocity $\beta$ and Lorentz factor $\gamma$ with the corrisponding constraint on the distance $l$ and decay width $\Gamma$. In all the calculation we set $m_X=16.85$ MeV.}
\label{table Atomki}
\end{table}

\newpage
\bibliographystyle{JHEP}
{\footnotesize
\bibliography{biblibimbi}}
\end{document}